\documentclass[aps,twocolumn,showpacs,prb,amsmath,amssymb,floatfix,superscriptaddress]{revtex4-1}

\usepackage{color}
\usepackage{tabu}
\usepackage{graphicx}
\usepackage{dcolumn}
\usepackage{bm}
\usepackage{array}
\usepackage{float}
\usepackage{supertabular}
\usepackage{longtable}
\usepackage{mathrsfs}
\usepackage{txfonts}
\usepackage{wasysym}
\usepackage{hyperref}

\begin{document}

\title{{Optimal diabatic dynamics of Majorana-based quantum gates}}

\author{Armin Rahmani}
\affiliation{Department of Physics and Astronomy and Advanced Materials Science and Engineering Center, Western Washington University, Bellingham, Washington 98225, USA}

\author{Babak Seradjeh}
\affiliation{Department of Physics, Indiana University, Bloomington, Indiana 47405, USA}

\author{Marcel Franz}
\affiliation{Department of Physics and Astronomy and Stewart Blusson Quantum Matter Institute, University of British Columbia, Vancouver, British Columbia, Canada V6T 1Z4}

\date{\today}


\begin{abstract}
In topological quantum computing, unitary operations on qubits are performed by adiabatic braiding of non-Abelian quasiparticles, such as Majorana zero modes, and are protected from local environmental perturbations. In the adiabatic regime, with timescales set by the inverse gap of the system, the errors can be made arbitrarily small by performing the process more slowly. To enhance the performance of quantum information processing with Majorana zero modes, we apply the theory of optimal control to the diabatic dynamics of Majorana-based qubits. While we sacrifice complete topological protection, we impose constraints on the optimal protocol to take advantage of the nonlocal nature of topological information and increase the robustness of our gates. By using the Pontryagin's maximum principle, we show that robust equivalent gates to perfect adiabatic braiding can be implemented in finite times through optimal pulses. In our implementation, modifications to the device Hamiltonian are avoided. Focusing on thermally isolated systems, we study the effects of calibration errors and external white and $1/f$ (pink) noise on Majorana-based gates. While a noise-induced \textit{antiadiabatic} behavior, where a slower process creates more diabatic excitations, prohibits indefinite enhancement of the robustness of the adiabatic scheme, our fast optimal protocols exhibit remarkable stability to noise and have the potential to significantly enhance the practical performance of Majorana-based information processing.

\end{abstract}
 
\maketitle
\section{Introduction}
Non-Abelian quasiparticles such as Majorana zero modes (MZMs) provide a promising platform for robust quantum information processing~\cite{kitaev_fault-tolerant_2003,nayak_non-abelian_2008}. Qubits are encoded in the fermion parities of MZM pairs nonlocally and are protected from local environmental perturbations. Quantum gates are implemented as unitary transformations in the degenerate ground-state manifold via adiabatic braiding of the MZM worldlines. There has been remarkable progress in realizing MZMs recently~\cite{oreg_helical_2010,lutchyn_majorana_2010,
alicea_new_2012,beenakker_2012,elliott_majorana_2015} and several experimental groups are working toward their braiding~\cite{mourik_signatures_2012,das_zero-bias_2012,churchill_superconductor_2013,rokhinson_fractional_2012,deng_anomalous_2012,finck_anomalous_2013,nadj-perge_observation_2014,aasen-milestones-2016}.

One of the challenges for an adiabatic scheme is the finite time of operations, causing inaccuracies in the unitary operations due to diabatic excitations~\cite{cheng_nonadiabatic_2011,karzig_boosting_2013,amorim_majorana_2014}. Other sources of error include quasiparticle poisoning~\cite{rainis_majorana_2012}, the on/off ratio of Coulomb coupling~\cite{vanheck_coulomb_2012}, and the information decay due to time-dependent perturbations~\cite{goldstein_decay_2011,schmidt_decoherence_2012}. While topological protection can certainly defend the system against many environmental perturbations, they do not make the system immune to errors like quasiparticle poisoning and high-frequency noise. These errors, in turn, limit the coherence time of the system, making it impossible to completely eliminate the diabatic excitations by sacrificing performance.

A few recent studies have addressed the diabatic excitations. One idea is adding counterdiabatic terms~\cite{berry_transitionless_2009,delcampo_shortcuts_2013} to the Hamiltonian of the system~\cite{karzig_shortcut_2015,zhang_shortcut_2015}. This scheme requires a reengineering of the devices and may pose experimental challenges. Another idea is to minimize the diabatic excitations by using smoother adiabatic protocols~\cite{knap_quick_braid_2016}. While improving the accuracy of the gates, this scheme still requires slower dynamics than the speed limit of the device.

A third approach is through the \textit{optimal control} of the quantum evolution~\cite{peirce_optimal_1988,palao_quantum_2002,kral_coherently_2007,
caneva_optila_2009,doria_optimal_2011,rahmani_optimal_2011,rahmani_quantum_2013}, in which we relax the requirement of remaining adiabatic during the evolution. Instead, we optimize the time dependence of the Hamiltonian parameters so as to generate the \emph{same} final state as the perfectly adiabatic dynamics. This approach relies only on optimizing pulse shapes and can be applied to existing experimental setups. It also realizes the characteristic speed limit of the device, resulting in the fastest possible information processing. Optimal control has been applied to the motion of one MZM along a one-dimensional wire~\cite{karzig_optimal_2015} but the full optimal creation of the same unitary gates as the adiabatic braiding remains an open question. 

In this paper, we solve the optimization problem exactly in the context of a simple effective model of MZM braiding. More generally, we address the following key questions: What is the speed limit for generating the same unitary evolution operator as the adiabatic braiding for two MZMs in our device? How robust are these operations to calibration errors and noisy pulses? By relaxing the constraint of adiabaticity during the entire process, we give up strict topological protection. Indeed a fully unconstrained optimal protocol, which only minimizes the difference between the evolution operator and a target unitary operator (corresponding to adiabatic braiding), would not utilize any of the topological features of the MZMs.

 In our optimal-control approach, we strike a balance between performance and robustness, by imposing constraints that can improve robustness against environmental perturbations, which utilize the nonlocal nature of information stored in pairs of MZMs. We then explicitly examine the effects of various errors on our gates and demonstrate remarkable practical advantages. For example, by calibrating our gates within a few percent of the desired values of control parameters, they outperform a topologically protected adiabatic gate and are faster by two orders of magnitude in the operation time. The shorter times of the gate operations defend our optimal protocols against decoherence sources like white noise and the experimentally important $1/f$ noise, allowing them to generate accurate unitary operations in much shorter times.
 
The remainder of this paper is organized as follows. In Sec.~\ref{sec:effective}, we review an effective low-energy model for the braiding of MZMs. In Sec.~\ref{sec:opt}, we first formulate the optimal-control problem and impose a constraint to increase the robustness of the protocol by making use of the nonlocal nature of information stored in pairs of MZMs. We briefly review the Pontryagin's minimum principle and use it to obtain the optimal protocol that generates the target adiabatic unitary operator exactly in a finite time. Section \ref{sec:error} is devoted to an in-depth study of the robustness of our optimal protocol. In Sec.~\ref{sec:error_model}, we first present a general noise model through a Taylor expansion of the control parameters ``seen'' by the system in terms of the control parameters imparted to the system experimentally. The leading error for a generic nontopological qubits is additive, while for topologically protected Majorana-based qubit the leading error is multiplicative.  In Sec.~\ref{sec:error_calib}, we examine the effects of systematic calibration error. In terms of a measure of distance between the unitary operators, the errors in optimal protocol grow linearly from zero. With multiplicative errors calibrated within 2\%, we find the optimal protocol can outperform an adiabatic protocol that is two orders of magnitude slower. In Secs.~\ref{sec:error_white} and \ref{sec:error_pink}, we consider random time-dependent errors, i.e., noise, in the control parameters. The effect of noise on adiabatic and optimal protocols is generally found to be very similar. For white noise, the fast optimal protocol outperforms all adiabatic protocols considered. For $1/f$ (pink) noise, an adiabatic gate that is 10 times slower starts to perform better at relatively large noise strength. We discuss a technique for correcting the errors caused by the limitations of our effective model in Sec.~\ref{sec:error_limi} and close the paper in Sec.~\ref{sec:conc} with a brief summary.

%------- Fig 1 --------
\begin{figure}[t]
\setlength{\unitlength}{1mm}
\begin{picture}(80,75)(0,0)
	\put(0,25){\large{(c)}}
	\put(2,0){\includegraphics[width=7.8cm]{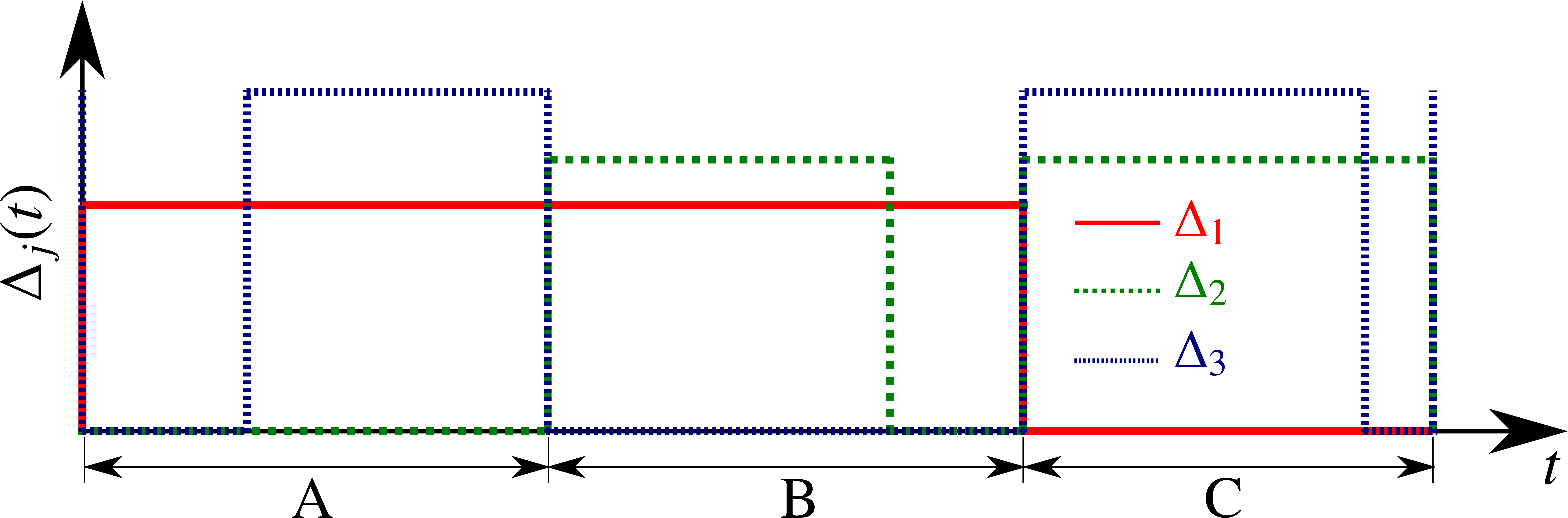}}
		\put(0,47){\large{(b)}}
	\put(2,25){\includegraphics[width=7.8cm]{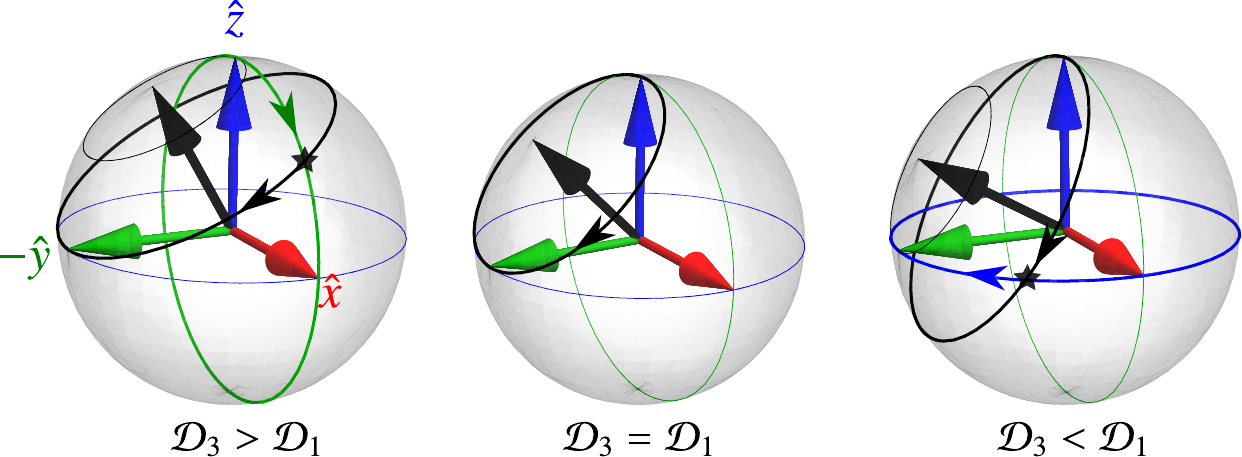}}
	\put(0,73){\large{(a)}}
	\put(2,55){\includegraphics[width=7.8cm]{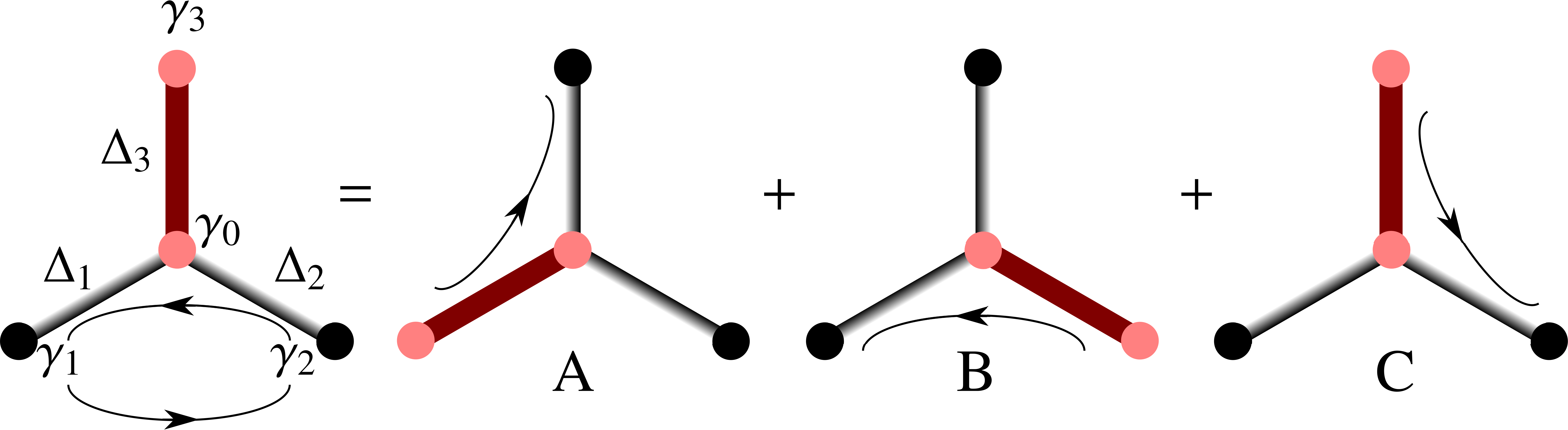}}
	\end{picture}
	\caption{(Color online) Optimal diabatic braiding of Majorana zero modes: (a) the three-step braiding scheme for exchanging $\gamma_1$ and $\gamma_2$; (b) the optimal diabatic trajectories in the Bloch sphere for step A (the black star indicates a switching from one axis of precession to another); and (c) the bang-bang optimal protocol for the entire process (with ${\cal D}_1<{\cal D}_2<{\cal D}_3$).}
	\label{fig:1}
\end{figure}
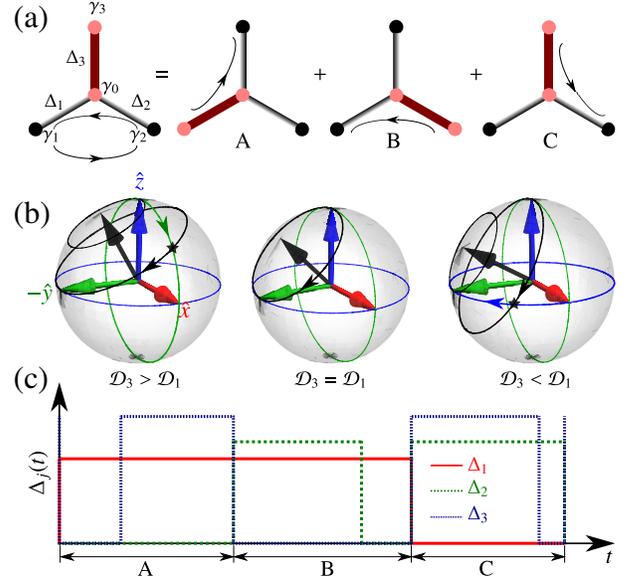
%------- Fig 1 --------

\section{Effective model of Majorana braiding}
\label{sec:effective}
We start from a minimal effective model of braiding, which is relevant to the current experimental efforts involving one-dimensional topological superconductors, e.g., in the top-transmon~\cite{ hassler_toptransmon_2011,
vanheck_coulomb_2012, hyart_flux_2013}. The Hamiltonian can be written in terms of four Majorana fermions as
%%%%%
\begin{equation}\label{eq:hamil0}
H(t)=i\gamma_0\sum_{j=1}^3\Delta_j(t)\gamma_j,
\end{equation}
%%%%%
where $\gamma_j=\gamma^\dagger_j$ and $\{\gamma_i,\gamma_j\}=2\delta_{ij}$. The coupling constant $\Delta_j$ represents the hybridization energy between $\gamma_j$ and $\gamma_0$. We assume that all $\Delta_j$ can be tuned as a function of time within a range $0\leqslant\Delta_j(t)\leqslant {\cal D}_j$. 
Defining two Dirac fermions $c=(\gamma_1+i\gamma_2)/2$ and $d=(\gamma_0+i\gamma_3)/2$,
we can write the Hamiltonian in the basis $\left(|0\rangle, d^\dagger c^\dagger |0\rangle,c^\dagger|0\rangle,d^\dagger|0\rangle\right)$ as a block-diagonal matrix,
\begin{equation}\label{eq:hamil}
{\cal H}_{\rm e(o)}=\Delta_1\sigma_y\mp\Delta_2\sigma_x-\Delta_3\sigma_z,
\end{equation}
%%%%%
where $\sigma_{x,y,z}$ are the Pauli matrices. The upper (lower) block ${\cal H}_{\rm e(o)}$ has even (odd) fermion parity.

The standard adiabatic scheme of braiding a MZM pair proceeds in three steps as depicted in Fig.~\ref{fig:1}. Starting with $\Delta_1=\Delta_2=0$ and $\Delta_3={\cal D}_3$, so that $\gamma_1$ and $\gamma_2$ are decoupled, we have two degenerate ground states, namely, $|0\rangle$ and $c^\dagger|0\rangle$, with opposite fermion parity. In each step, we adiabatically turn on one coupling to its maximum value and turn off another to zero. At the end of the three steps, we return to the initial Hamiltonian, generating a unitary transformation $U=\exp\left(\pi\gamma_2\gamma_1/4\right)$ in the ground-state manifold. In the $(|0\rangle,c^\dagger|0\rangle )$ basis, up to an unimportant overall phase, we can write $U={\rm diag}(1,i)$, hereafter referred to as the target unitary.

Our goal is to generate (up to a phase) the target unitary via diabatic evolution of $\Delta_j(t)$ in a finite total time $\tau$. The permissible diabatic protocols are bounded functions $0<\Delta_j(t)<{\cal D}_j$ over the time interval $0<t<\tau$. The shortest time, $\tau^*$, for which it is possible to generate the target unitary with a permissible protocol sets the speed limit of the device.

\section{Optimal control approach}
\label{sec:opt}
The most general diabatic protocols allow for the hybridization of all the MZMs, which destroys the topological protection.
As discussed in Refs.~\cite{goldstein_decay_2011,schmidt_decoherence_2012}, adiabatic braiding is not protected against perpetual dynamical perturbations specially if they have high-frequency components. External noise can also result in an antiadiabatic behavior~\cite{dutta} for very slow ramps (see also our Fig.~\ref{fig:3} and its discussion). Moreover, the long time scales required to create accurate gates with the adiabatic evolution, under which the operation enjoys topological protection, may overshoot the coherence time of the system, which is limited by, e.g., quasiparticle poisoning. Topological protection, however, implies robustness to a wide range of local perturbations and, in particular, static calibration errors. One approach would be to altogether abandon the benefits of information nonlocality and simply optimize $0<\Delta_j(t)<{\cal D}_j$ to minimize the difference of the evolution operator and the adiabatic transformation $U$. However, this requires extreme fine-tuning and exposes the gate operation to an array of unwanted perturbations. Instead, we take a balanced approach that utilizes the nonlocal nature of the qubits while improving its operation speed by orders of magnitude.

We constrain the optimal dynamics to track the same three-step dynamics as in the adiabatic scheme, without requiring adiabaticity \textit{during} each step.  For example, throughout step A, we keep $\Delta_2=0$ and change $\Delta_1$ and $\Delta_3$ in their permissible range. Therefore, $\gamma_2$ remains decoupled and the parity of the $c$ fermion cannot be accessed by local environmental perturbations. As the total parity is conserved, the parity of the $d$ fermion is also locally inaccessible \emph{despite} the generation of diabatic excitations during the evolution. Similarly, in step B (C), we keep $\Delta_3=0$ ($\Delta_1=0$) and decouple $\gamma_3$ ($\gamma_1$). This way, step A is protected from local environmental perturbations. If we execute step A perfectly, then we have a decoupled MZM at the beginning and during step B, and step B will be protected as well. The sacrifice to topological protection originates from possible inaccuracies in step A, which can propagate to the next steps. By design, at the end of each step (but not during), the state of the system is optimized to mimic a fully adiabatic evolution.

Focusing on step A with $\Delta_2=0$, we have ${\cal H}_{\rm e}={\cal H}_{\rm o}=\Delta_1\sigma_y-\Delta_3\sigma_z$. Let us concentrate on one parity sector.  The initial state is the ground state for $\Delta_1=0$ and $\Delta_3>0$, i.e., the eigenstate of $\sigma_z$ with eigenvalue $+1$,  $|+z\rangle$. The target state at the end of step A is the ground state for $\Delta_1>0$ and $\Delta_3=0$, i.e., $|-y\rangle$. Denoting the total time with $\tau_A$, we minimize the following functional of $\Delta_{1,3}(t)$:
%%%%%
\begin{equation}\label{eq:F_A}
{\cal F}_A=1-|\langle -y|\mathrm{T}e^{{-i\int_0^{\tau_A}\left[\Delta_1(t)\sigma_y-\Delta_3(t)\sigma_z\right]dt}}|+z\rangle|^2,
\end{equation}
where $\mathrm{T}$ indicates time ordering.
For a given $\tau_A$, the optimal protocol yields the smallest possible ${\cal F}_A$. As we increase $\tau_A$, this minimal ${\cal F}_A$ decreases and eventually vanishes for a critical time $\tau_{A}^*$, where the target state is prepared exactly. 

To compute $\tau_{A}^*$ and the corresponding optimal protocol, we use Pontryagin's maximum principle~\cite{pontryagin_mathematical_1987}. 
The principle states that for dynamical variables ${\bf x}=(x_1, x_2, \cdots x_n)$ and control functions $\boldsymbol\Delta=(\Delta_1, \Delta_2, \cdots \Delta_m)$, evolving with the equations of motion ${d \over dt}{\bf x}={\bf f}({\bf x},\boldsymbol\Delta)$ from a given initial conditions ${\bf x}(0)$ to a final set ${\bf x}(\tau)$, the optimal controls, $\boldsymbol\Delta^*$, which minimize a cost function ${\cal F}[{\bf x}(\tau)]$ (any function of the final values of the dynamical variables), satisfy
%%%%%
\begin{equation}\label{eq:pontry}
{\bf p}^*\cdot{\bf f}({\bf x}^*,\boldsymbol\Delta^*)=\min_{\boldsymbol\Delta}\left[{\bf p}^*\cdot {\bf f}({\bf x}^*,\boldsymbol\Delta)\right],
\end{equation}
%%%%%
where $\bf p$ are conjugate dynamical variables with equations of motion,
\begin{equation}
{d \over dt}{\bf p}=-{\bf p}\cdot\frac{\partial}{\partial {\bf x}}\:{\bf f}({\bf x},\boldsymbol\Delta),
\end{equation}
and $\bf x^*$ and $\bf p^*$ are optimal trajectories corresponding to $\boldsymbol\Delta^*$. Furthermore, the boundary condition for $\bf p$ is set by the cost function as 
$
\bf p(\tau)=\frac{\partial}{\partial {\bf x}}{\cal F}[{\bf x}(\tau)].
$

As a consequence of Eq.~\eqref{eq:pontry}, 
when $\bf f$ [and consequently ${\bf p}^*\cdot{\bf f}({\bf x}^*,\boldsymbol\Delta)$] 
are linear functions of the controls $\boldsymbol\Delta$, the optimal protocols are ``bang-bang'': each of the control functions $\boldsymbol\Delta$ attain either its minimum or maximum allowed value at any given time (unless the coefficient of a component of $\boldsymbol\Delta$ identically vanishes over a finite interval~\footnote{This special scenario does not occur in our system.}).

In the problem at hand, the real and imaginary parts of the wave function serve as dynamical variables, with equations of motion given by the Schr\"odinger equation, which is indeed linear in the controls $\Delta_j$. Also, the cost function Eq.~\eqref{eq:F_A} depends only on the final wave function. Therefore, of all the permissible functions $\Delta_{1,3}$, the optimal protocols are discontinuous functions that either vanish or attain their maximum allowed value ${\cal D}_{1,3}$ at any given time. We cannot have $\Delta_1=\Delta_3=0$ for optimal control since then the Hamiltonian would vanish and the state would not evolve. Thus, the optimal protocol consists of a sequence of potentially three types of Hamiltonians with sudden switchings between them. Due to the mapping of the Hamiltonian for each parity sector to a spin-$1/2$, we can visualize the dynamics on the Bloch sphere.  If only $\Delta_1$  ($\Delta_3$) is turned on, the quantum state precesses around the $y$ ($z$) axis in the Bloch sphere. If both couplings are turned on, it precesses around an intermediate axis shown in black in Fig.~\ref{fig:1}(b). 

We now identify the minimal path corresponding to the critical time $\tau_{A}^*$. This simultaneously determines the optimal protocol and the minimum required time for an exact state transformation. As seen in Fig.~\ref{fig:1}(b), in the special case with ${\cal D}_1={\cal D}_3={\cal D}$, the protocol is extremely simple. We turn on both couplings to their maximum and a single precession prepares the target state exactly in a time $\tau_{A}^*={\pi / \left(2\sqrt{2} {\cal D}\right)}$. In the general case, we only need one switching during the process as shown in Fig.~\ref{fig:1}(b). The general form of the optimal protocol in a step that transfers a MZM from leg $a$ to leg $b$ is as follows. If ${\cal D}_a\leqslant {\cal D}_b$, we first switch on $\Delta_a={\cal D}_a$ while keeping $\Delta_b=0$, wait for a time ${1\over 2{\cal D}_a}\cos^{-1}\left({\cal D}_a/{\cal D}_b\right)$, and then switch on $\Delta_b={\cal D}_b$ for a time ${1\over 2\sqrt{({\cal D}_a)^2+({\cal D}_b)^2}}\cos^{-1}\left[-\left({\cal D}_a/{\cal D}_b\right)^2\right]$. For ${\cal D}_a\geqslant{\cal D}_b$, due to time-reversal symmetry, the process is the same in reverse. An example of such optimal protocol, combining all three steps, is shown in Fig.~\ref{fig:1}(c). While in steps B and C, ${\cal H}_{\rm e}\neq {\cal H}_{\rm o}$, it turns out that for both blocks, the initial state is transformed to the target state by the same protocol.

We now explicitly compute the non-Abelian unitary operator generated by the optimal protocols above. Without loss of generality, we consider the case ${\cal D}_1\leqslant {\cal D}_2\leqslant {\cal D}_3$. Using the notation $s_{ij},d_{ij}\equiv \sqrt{{\cal D}_i\pm{\cal D}_j}$, we can write
\begin{equation}
\begin{split}
U_{\rm e(o)}={1\over 8{\cal D}^3_3{\cal D}^{3/2}_2}
&\left(s_{32}\pm id_{32}\sigma_x\right)\left(s_{32}d_{32}\pm i{\cal D}_2\sigma_x+i{\cal D}_3\sigma_z\right)\\
&\times\left(s_{21}- id_{21}\sigma_y\right)\left(s_{21}d_{21}\pm i{\cal D}_2\sigma_x-i{\cal D}_1\sigma_y\right)\\
&\times\left(s_{31}d_{31}- i{\cal D}_1\sigma_y+i{\cal D}_3\sigma_z\right)\left(s_{31}- id_{31}\sigma_y\right).
\end{split}
\end{equation}
Despite the complexity of the above unitaries, it can be verified that the evolution operator [generated by the optimal protocols as in Fig. 1(c)], projected to the ground state manifold, $|0\rangle$ and $c^\dagger|0\rangle$), i.e., ${\rm diag}(U^{11}_{\rm e},U^{11}_{\rm 0})$, equals the target unitary $U$ up to an overall phase.
%An explicit computation of the non-Abelian unitary operator produced by the optimal protocols above (see Fig. 1c)  shows that the evolution operator projected to the ground-state manifold, $|0\rangle$ and $c^\dagger|0\rangle$, indeed equals the target unitary $U$ up to an overall phase.

% A tedious but straightforward calculation 
\section{errors and robustness}
\label{sec:error}
\subsection{Error model} 
\label{sec:error_model}
Since our optimal bang-bang protocols are fine-tuned to the parameters of the device, one should naturally wonder how robust the process is. We consider two types of errors: (i) \emph{calibration errors} that arise from the absence of precise knowledge about the actual effective Hamiltonian parameters; (ii) \emph{random errors} due to the imperfect control over the external knobs, e.g., gate voltages, which make the parameters noisy. The errors of type (i) are systematic and can be minimized by careful calibration. The errors of type (ii), on the other hand, generate a different final state every time the experiment is run. We demonstrate that even in the presence of these errors, our scheme presents advantages over the adiabatic methods.

We begin by modeling the errors. Generically, attempting to tune a coupling to $\Delta_j$ imparts to the system an effective $\Delta^S_j$. The error can be expanded (at any point in time) in $\Delta_j(t)/{\cal D}_j<1$ as
\begin{equation}\label{eq:calib}
\Delta^S_j\approx\Delta_j+{\cal D}_j\left[\epsilon_{j0} +\epsilon_{j1} \left(\Delta_j/{\cal D}_j\right)+\epsilon_{j2}  \left(\Delta_j/{\cal D}_j\right)^2+\dots\right].
\end{equation}
Calibration errors are characterized by time-independent $\epsilon_{jn}$~\cite{karzig_geometric_2016}, whereas random errors are modeled by noisy $\epsilon_{jn}$. Here we focus on Gaussian white noise with
second moment 
\begin{equation}\label{eq:moment}
R^{jnj'n'}_{\rm white}(t-t')=\overline{\epsilon_{jn}(t)\epsilon_{j'n'}(t')}=W_{jn}^2\delta_{jj'}\delta_{nn'}\delta(t-t')
\end{equation}
 and noise strength $W_{jn}$ as well as $1/f$ (pink) noise, which is expected to be the dominant source of noise in experiments. For white noise, the spectral noise density defined as the Fourier transform of the correlation function, i.e., $S(\omega)\equiv\int_{-\infty}^{\infty}R(t)e^{-i\omega t}dt $ is a constant, while for pink noise it decays as 
\begin{equation}\label{eq:pink}
S^{jnj'n'}_{\rm pink}(\omega)\sim 
\delta_{jj'}\delta_{nn'}{1\over \omega}.
\end{equation}

In the case of white noise, we compute the noise-averaged density matrix through a numerically exact solution of a Lindblad-type master equation. Due to the correlations in pink noise, the noise-averaged density matrix evolves with an integral equation that is difficult to solve. We therefore resort to discrete Langevin-type numerical simulation, where we generate many discrete realizations of noise, evolve the system for each with the Schr\"odinger equation, and average the density matrices at the end. 

For nontopological qubits, the leading error is the additive error $\epsilon_{j0}$. However, for topological qubits, e.g., in the top-transmon, the coupling is generated by the overlap of Majorana wave functions; so in the limit $\Delta_j\to0$ all errors are exponentially small~\cite{hassler_toptransmon_2011,
vanheck_coulomb_2012, hyart_flux_2013}. Thus, the additive error is irrelevant for Majorana-based topological qubits and the leading error is the multiplicative $\epsilon_{j1}$. In the following, we present results for both additive and multiplicative errors. However, only the multiplicative error is relevant to topological qubits.

\subsection{Calibration errors}
\label{sec:error_calib}
We first discuss  calibration errors. Evolving the system with a given protocol generates an evolution operator $S$ in the ground-state manifold. We quantify the deviation from the target unitary $U$ by the distance~\cite{zhang}
\begin{equation}
{\cal E}(S,U)\equiv \sqrt{1-|{\rm tr}(S^\dagger U)/{\rm tr}({\openone})|^2},
\end{equation}
which is independent of the initial state. The target unitary $U={\rm diag}(1,i)$ lives in the ground-state manifold and $S$ is the projection of the full evolution operator to this manifold. Although $S$ may not be unitary after this projection, ${\cal E}(S,U)$ still provides a sensible measure of distance.

%------ Fig 2 -------
\begin{figure}[t]
	\includegraphics[width=8.3cm]{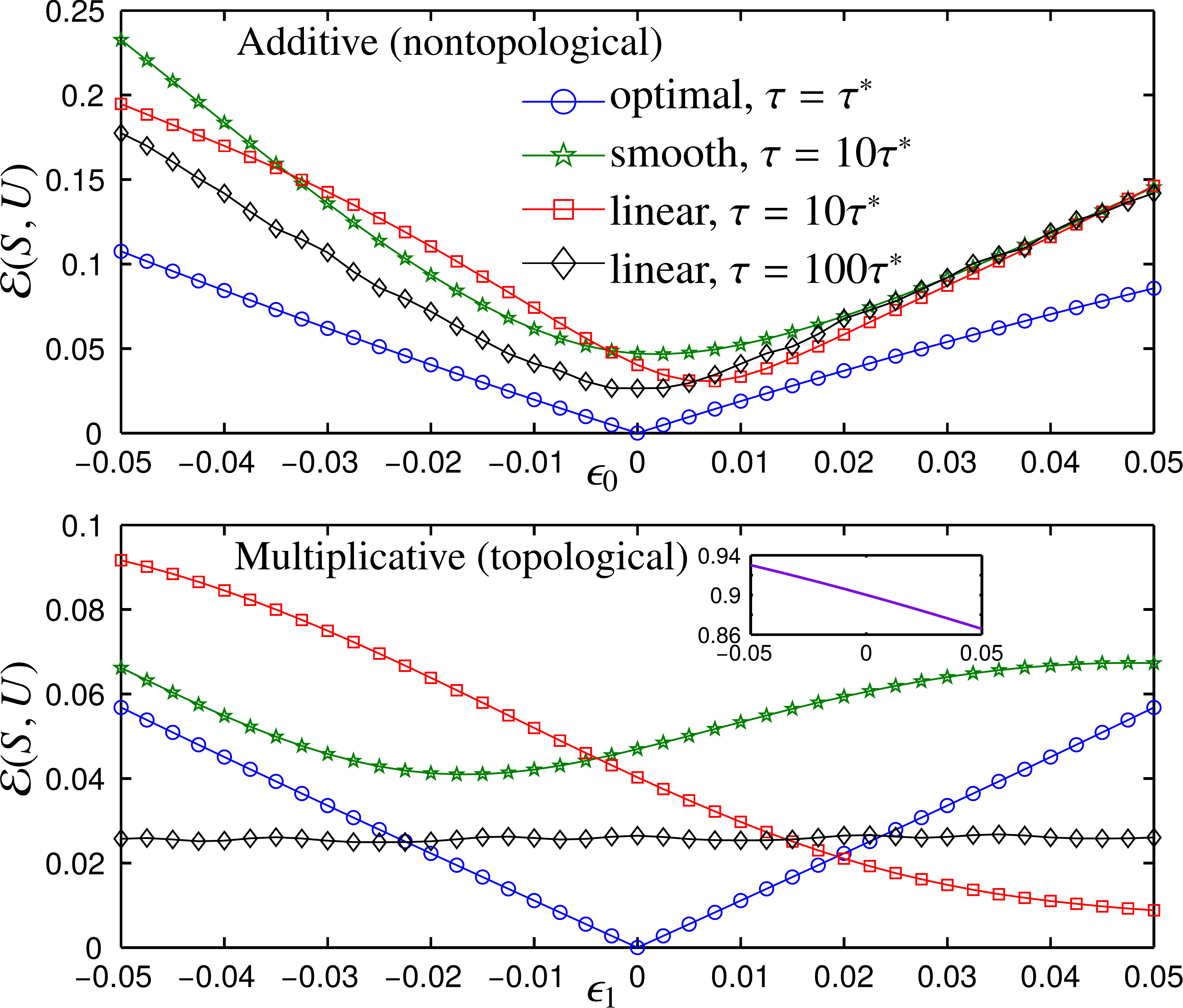}
	\caption{(Color online) The effects of calibration errors. The distance ${\cal E}(S,U)$ of the actual evolution operator $S$ to the target unitary $U$ as a function of additive $\epsilon_0$ (top) and multiplicative $\epsilon_1$ (bottom) calibration errors for optimal diabatic protocol at $\tau=\tau^*$ as well as the linear and smooth adiabatic protocol at $\tau=10\tau^*$ and $\tau=100\tau^*$. The inset shows ${\cal E}(S,U)$ vs. $\epsilon_1$ for the linear protocol for $\tau=\tau^*$.}
	\label{fig:2}
\end{figure}
%------ Fig 2 -------

For concreteness, we focus on the case ${\cal D}_1={\cal D}_2={\cal D}_3=1$, where the optimal protocols are simple. In the adiabatic schemes, each step is done in a time $T=\tau/3$. We consider two types of adiabatic protocols: linear switches $\Delta_j^{\rm on}(t)=t/T$ and $\Delta_j^{\rm off}(t)=1-t/T$; and smooth switches $\Delta_j^{\rm on}(t)=\sin^2\left({\pi}t/2T\right)$ and $\Delta_j^{\rm off}(t)=\cos^2\left({\pi}t/2T\right)$ with vanishing slopes at the boundaries of the steps. Here $0<t<T$ is measured from the beginning of each step. For all of these protocols (optimal, linear, and smooth), the evolution of the system is governed through $\Delta^S_j(t)$ as in Eq.~\eqref{eq:calib} to leading order (additive $\epsilon_{j0}$ and multiplicative $\epsilon_{j1}$, respectively, for generic and topological qubits). For simplicity we take $\epsilon_{jn}=\epsilon_n$ independent of $j$. The optimal protocol for $\epsilon_k=0$ generates the target unitary $U$ exactly in a time $\tau^*=3\tau_A^*=3\pi/\left(2\sqrt{2}\right)$. The linear and smooth protocols over the same time are completely nonadiabatic (see the inset of  Fig.~\ref{fig:2}). Therefore, instead of a comparison over the same time, we compare the optimal protocol with adiabatic protocols that are at least one order of magnitude slower.

In Fig.~\ref{fig:2}, we show the error ${\cal E}(S,U)$ as a function of additive and multiplicative calibration error. As expected, there are no advantages for an adiabatic protocol in the nontopological case of additive error. On the other hand, topological protection gives rise to robust adiabatic protocols when errors are multiplicative.
For timescales that are an order of magnitude larger, the adiabatic methods are sensitive to the pulse shape and the calibration error $\epsilon_1$. At time scales that are two orders of magnitude larger, the adiabatic method becomes insensitive to $\epsilon_1$  and starts to outperform the optimal protocol for errors larger than $2\%$ (note that $\epsilon_1$ is dimensionless). Upon further increasing $\tau$, the robust error of the adiabatic method decreases further. However, the lower speed of the operation is limited by coherence times and it is impractical to keep slowing down the process. The fast optimal protocol, which has a fixed short time $\tau^*$, can perform better than any adiabatic gate, upon improved calibration. As seen in Fig.~\ref{fig:2}, the error ${\cal E}(S,U)$ for the optimal protocol has a linear dependence on $\epsilon_1$. We also note that when the error is multiplicative rather than additive the optimal protocol also performs better.

%------ Fig 3 -------
\begin{figure}[t]
	\includegraphics[width=8cm]{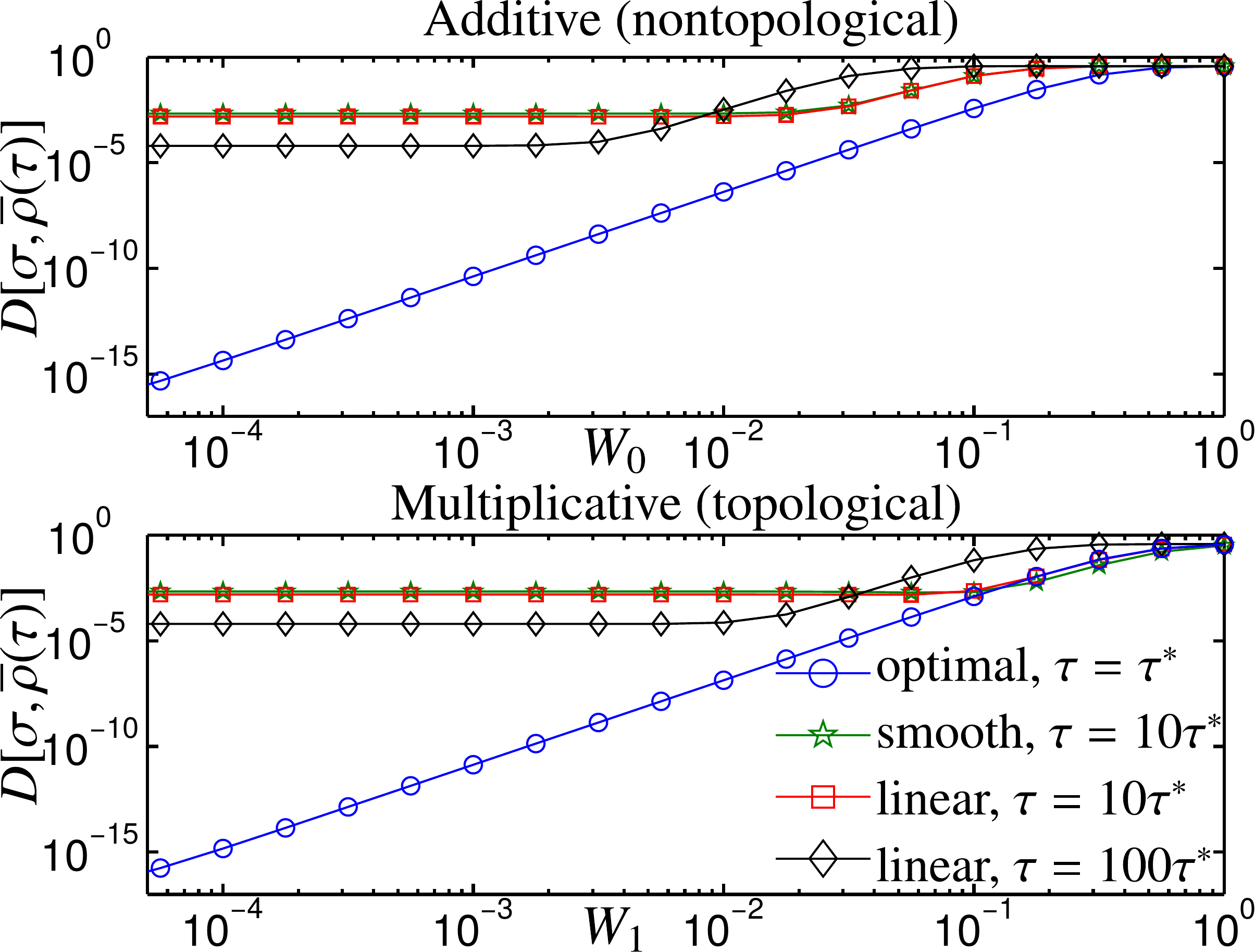}
	\caption{(Color online) The effects of random noise. The trace distance between the final and the target density matrices for an equal-weight initial superposition of the ground states as a function of the noise strength, $W$, for optimal diabatic, and linear and smooth adiabatic protocols.
	\label{fig:3}}
\end{figure}
%------ Fig 3 -------

\subsection{Random white noise}
\label{sec:error_white}
We now turn to the noisy couplings. While systematic errors can be potentially corrected by careful calibration, random time-dependent errors pose a greater challenge to both the adiabatic and optimal gates. We start by quantifying the errors due to noise.
Noise averaging is essential when dealing with random protocols. Direct averaging of the unitaries, however, creates artificial dephasing due to the unimportant overall $U(1)$ factors. Thus, we need a different cost function. We choose to work with the noise-averaged density matrix, $\overline{\rho}$. We start from a particular superposition of the ground states as the initial state, $|\psi_0\rangle=\frac1{\sqrt 2}\left(|0\rangle+|1\rangle\right)$, where $|1\rangle \equiv c^\dagger |0\rangle$, yielding the initial density matrix $\rho_0=\frac12\left(|0\rangle \langle 0|+|0\rangle \langle 1|+|1\rangle \langle 0|+|1\rangle \langle 1|\right)$, which is then evolved and averaged over noise to obtain $\overline{\rho}(\tau)$, by solving the master equation~\cite{pilcher_heating_2013, rahmani_dynamics_2015,dutta},
%%%%%
\begin{equation}\label{eq:master} 
\partial_t{\overline{\rho}}=-i[H,\overline{\rho}] 
-{1\over 2}\sum_{n,j}W_{jn}^2{\cal D}_j^{2(1-n)}\Delta_j^{2n}\left[\left[\overline{\rho},i\gamma_0\gamma_j\right],i\gamma_0\gamma_j\right].
\end{equation}
%%%%%
%with the initial condition $\overline{\rho}(0)=\rho_0$.
%~\footnote{ In our 4-dimensional basis [see Eq.~\eqref{eq:hamil}], we find ${d\over dt}\overline{\rho}(t)=-i\sum_j[\Delta_j(t)O_j,\overline{\rho}(t)]+W^2\sum_j O_j \overline{\rho}(t) O_j-{3}W^2\overline{\rho}(t)$, where the matrices $O_j$ are given by $O_{1}=\openone\otimes \sigma_{y}$, $O_{2}=- \sigma_{z}\otimes \sigma_x$, and $O_{3}=-\openone\otimes \sigma_{z}$. }.
The target state $U|\psi_0\rangle$ yields the target density matrix $\sigma=\frac12\left(|0\rangle \langle 0|-i|0\rangle \langle 1|+i|1\rangle \langle 0|+|1\rangle \langle 1|\right)$. We then quantify the error by the trace distance,
\begin{equation}
D[\sigma,\overline{\rho}(\tau)]\equiv {1\over 2}{\rm tr}\sqrt{\left[\sigma-\overline{\rho}(\tau)\right]^2}.
\end{equation} We consider the leading order with$j$-independent noise, where only $W_{j0}$ and $W_{j1}$ are nonzero, respectively, for the nontopological and topological qubits.

Numerically solving for $\overline{\rho}(\tau)$ and computing the trace distance for the optimal as well as linear and smooth adiabatic protocols up to $\tau=100\tau^*$ indicates that the optimal protocol generally outperforms the adiabatic protocols for both additive and multiplicative noise. For $W=0$, the optimal protocol produces a vanishing trace distance, which then grows as $W^2$, while remaining much smaller than the trace distance corresponding to the adiabatic schemes before reaching saturation. Only for $\tau=10\tau^*$ the smooth protocol performs slightly better than the optimal protocol for multiplicative noise (as seen in a barely noticeable crossing of the green and blue curves in the bottom panel of Fig. \ref{fig:3}). However, this occurs in the regime of relatively large $D[\sigma,\overline{\rho}(\tau)]>10^{-3}$ and large $W_1>10^{-1}$. Interestingly, there is a crossing of adiabatic curves with $\tau=10\tau^*$ and $\tau=100\tau^*$ in Fig.~\ref{fig:3} for both additive and multiplicative noise, beyond which increasing the time scales of the adiabatic protocols reduces their robustness. This \textit{antiadabatic} behavior appears analogous to the anti-Kibble-Zurek behavior~\cite{dutta}.

We comment that in real experiments, a weakly coupled bath is always present, which is neglected in our analysis. If the bath decoheres the system, both adiabatic and optimal schemes fail (as quantum coherence is necessary for quantum information processing).

%; in our case, it is a limitation of adiabatic processes, which further motivates the use of the optimal protocol.
\subsection{Pink, $1/f$ noise}
\label{sec:error_pink}
White noise allows for numerically exact calculations through the solution of a deterministic differential master equation [see Eq.~\eqref{eq:master}] . This limit is relevant under more general conditions than those suggested by its precise mathematical definition, e.g., to the ubiquitous Ornstein-Uhlenbeck process, where the correlations of noise in the time domain decay exponentially. Intuitively, exponentially decaying correlations can be safely cut off after a characteristic correlation time, recovering the white-noise predictions upon temporal rescaling~\cite{Dalessio}. However, we expect the noise spectra in real experiments to have a $1/f$ frequency dependence~\cite{hassler_toptransmon_2011}.

Before a quantitative analysis of $1/f$ noise, we comment that qualitative similarities between the effects of white noise and other types of colored noise are expected.
Noise introduces a rate for the deposition of excess energy, which can be understood by viewing it as a sequence of small quantum quenches. Each quench deposits some energy into the system without a strong dependence on the deterministic part of the Hamiltonian $H_0(t)$. Whether there are correlations between these quenches (colored noise) or they are completely uncorrelated (white noise) should not qualitatively alter this generic effect. This does not imply, however, that the spectral density of the noise is unimportant. An extreme case is a noise spectrum localized on certain frequencies, which are either resonant or lie outside the bandwidth of the system, respectively, enhancing or suppressing the absorption of energy by the system. Such localized noise spectra are not common in experiment.

Unlike white noise, the temporal correlations of $1/f$ noise do not allow us to compute the noise-averaged density matrix by solving a single deterministic differential equation. We therefore take a brute-force approach of direct Langevin-type numerical simulations, where we use the method of Ref.~[\onlinecite{Kasdin}] to generate the discrete noise signal. This method applies to $1/f^\alpha$ noise spectrum, with $\alpha=0$ ($\alpha=1$) corresponding to white (pink) noise. In this section, we only present the results of Langevin-type simulations for the $1/f$ noise with $\alpha=1$. As a benchmark, we have checked, however, that the method indeed reproduces the same results as Sec.~\ref{sec:error_white} for the white-noise case.

For the numerical simulations, we first divide the total time of the process into $N$ intervals of duration $\Delta t=\tau/N$.
We only consider the multiplicative noise in this section (which is relevant to topological qubits) and keep the simplifying assumption ${\cal D}_j=1$. In terms of a correlated discrete signal $x_m$, the discretized noisy coupling constants then take the form
\begin{equation}
\Delta_j^S(t)=\Delta_j(t)\left(1+{Wx_m\over \Delta t^{1/2-\alpha/2} }\right), \quad (m-1)\Delta t<t<m\Delta t, 
\end{equation}
with $m=1 \dots N$.
The discrete noise signal $x_m$ is, in turn, generated from an uncorrelated zero-mean Guassian signal $w_m$ with a standard deviation of unity ($\overline{w_m}=0$ and $\overline{w_m w_n}=\delta _{mn}$) by using an autoregression model of finite order $M$ that relates the two signals through~\cite{Kasdin}:
\begin{equation}
\sum_{m=0}^M{\Gamma(m -\alpha/2)\over m!\Gamma( -\alpha/2)}x_{n-m}=w_n.
\end{equation}
Here, $\Gamma$ represents the gamma function. 

We are interested in the limit of $\Delta t\to 0$ so that our discrete simulations can provide a reasonable approximation to the continuous process. To this end,  we first fix $\Delta t$ and generate enough realizations to achieve convergence (within acceptable error bars) for the final numerically computed error $D[\sigma,\overline{\rho}(\tau)]$. We then increase $N$ so these realization-averaged errors also converge in $\Delta t$. Achieving perfect convergence in these calculations is time consuming specially for longer times and larger strengths of noise. Nevertheless, by analyzing $20000$ realizations and five different values of $\Delta t=\tau^*/N$ for $N=30,60,90,120,150$ (for the adiabatic processes with longer $\tau$, we increased $N$ to have the same values of $\Delta t$), we were able to significantly reduce the error bars.

\begin{figure}[]
	\includegraphics[width=8cm]{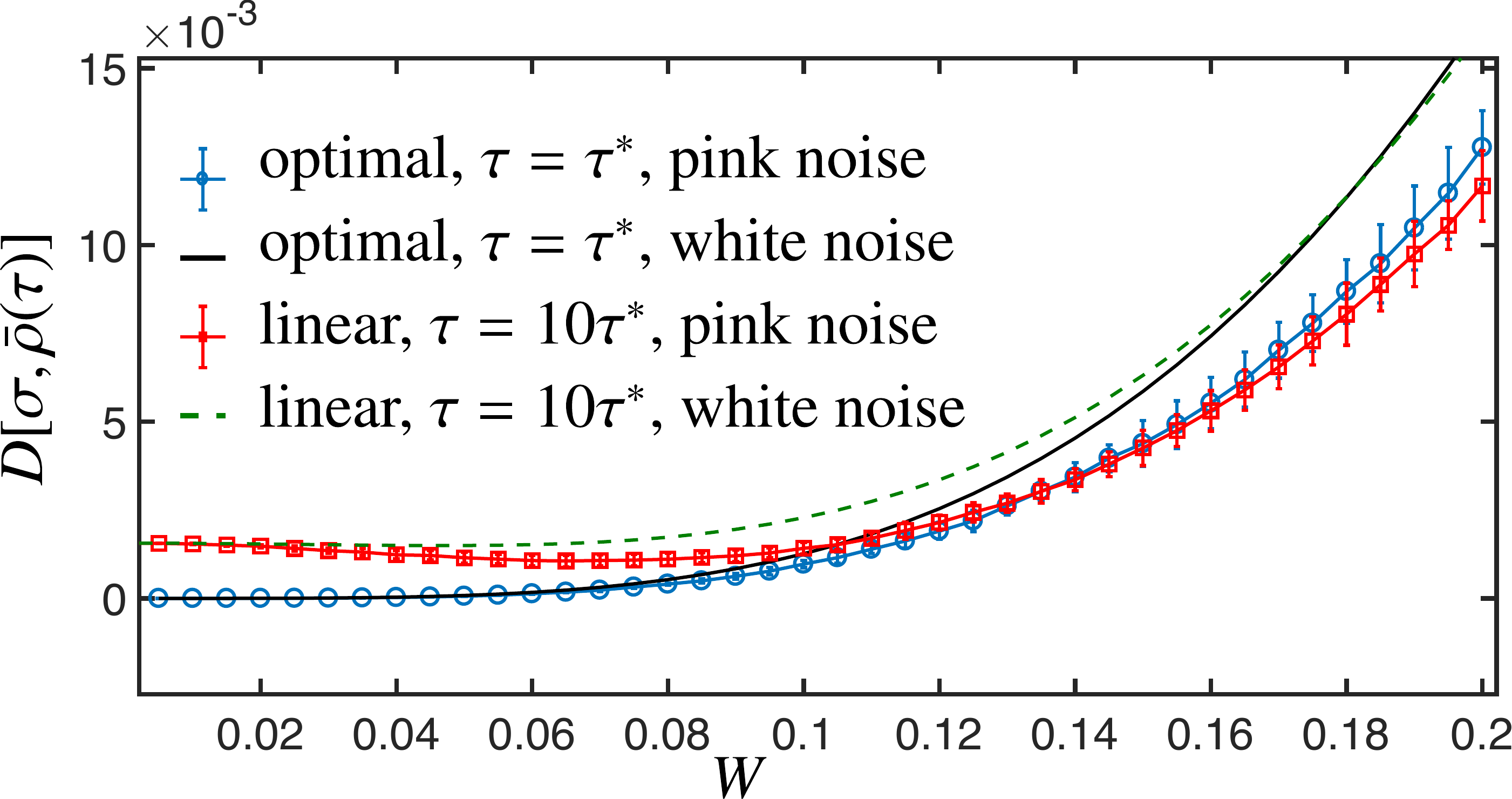}
	\caption{\label{fig:4}(Color online) 
	The effects of pink noise on both the optimal bang-bang and a linear adiabatic protocol that is one or two orders of magnitude slower. The white-noise data from Fig.~\ref{fig:3} is also replotted for easy comparison.}
\end{figure}

The results are shown in Fig. \ref{fig:4}. As expected, both the bang-bang optimal protocol and the linear adiabatic protocol are affected by the $1/f$ noise in a qualitatively similar manner to white noise. Due to the suppression of high-frequency modes, the $1/f$ noise has a milder effect than white noise on both of these protocols. These numerical results support our qualitative picture of the effects of noise. The advantages of the optimal protocol survive  in the noise regime, where the $D[\sigma,\overline{\rho}(\tau)]$ errors are small. Interestingly, the antiadiabatic behavior, where a slower adiabatic process, created more diabatic excitations, also occurs for the $1/f$ noise. In particular, an adiabatic process with $\tau=100\tau^*$ begins to underperform a process with $\tau=10\tau^*$ at $W\sim 0.04$. While this effect might appear counterintuitive, it naturally results from the accumulation of noise-induced excitations over a longer period. The antiadiabatic behavior further motivates the use of the optimal protocol. 

\subsection{Correcting the errors due to the limitations of the model}
\label{sec:error_limi}
Our results are obtained in the context of the effective model Eq.~\eqref{eq:hamil0}, which is written in terms of low-energy degrees of freedom and has an infinite gap to higher excitations. The optimal protocol involves sharp sudden quenches, which, in a more realistic model with a finite excitation gap, may cause high-energy excitations. In this section, we fix this issue by introducing an alternative cost function (for each of the three steps of the protocol) that penalizes sharp transitions and yields continuous optimal protocols that only take slightly longer than $\tau^*$.

We introduce a modified optimal-control problem for each of the three steps of the dynamics, where, e.g., in step A, we minimize 
\begin{equation}\label{eq:G_A}
{\cal G}_A(\lambda)=(1-\lambda){\cal F}_A+\lambda \int_0^{\tau_A}\left[\left(\frac{d\Delta_1}{dt}\right)^2+\left(\frac{d\Delta_3}{dt}\right)^2\right]dt,
\end{equation}
%\begin{equation}\label{eq:G_A}
%{\cal G}_A(\lambda)=(1-\lambda){\cal F}_A+\lambda \int_0^{\tau_A}\left[\left(\frac{d\Delta_1}{dt}\right)^2+\left(\frac{d\Delta_3}{dt}\right)^2\right]dt,
%\end{equation}
instead of minimizing, e..g., ${\cal F}_{A}$, with the constraints $\Delta_1(0)=0$, $\Delta_3(0)={\cal D}_3$, $\Delta_3(\tau_A)=0$, and $\Delta_1(\tau_A)={\cal D}_1$ (and similarly for steps B and C). The second term penalizes large derivatives in the protocol, turning the sudden jumps into continuous ramps. The weight $0\leqslant\lambda\leqslant 1$ sets the time scale of the ramps from 0 for $\lambda=0$ to the total time of the step for $\lambda=1$, in which case we get a simple linear protocol from Euler-Lagrange minimization of ${\cal G}_A(1)$.
\begin{figure}[t]
	\includegraphics[width=7.5cm]{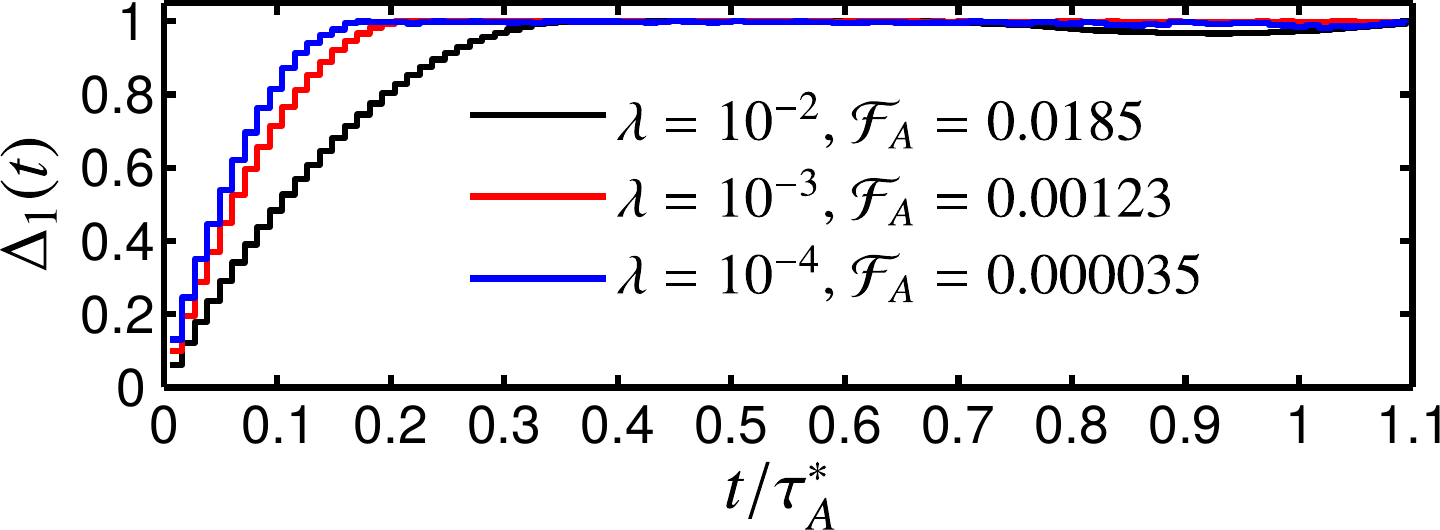}
	\caption{(Color online) The continuous protocol. The plot shows protocols for ramping up $\Delta_1$ in step A with ${\cal D}_1={\cal D}_3=1$ obtained from Monte Carlo optimization by allowing a total time $\tau=1.1\tau^*$, as well as the corresponding distance ${\cal F}_A$ to the target state, for various weights $\lambda$. 
	\label{fig:5}}
\end{figure}

While the Pontryagin's formalism can also shed light on optimal-control problems with a trajectory-dependent cost function as in Eq.~\eqref{eq:G_A}, an analytical solution of the constrained problem is challenging. We therefore use direct numerical minimization. Approximating a general protocol with a piece-wise constant protocol with $N=100$ steps, we perform Monte Carlo simulations over the shape of the protocols to minimize ${\cal G}_A(\lambda)$ for several values of $\lambda$ over a total time $\tau_A=1.1\tau_A^*$. The results for ramping up $\Delta_1$ are shown in Fig.~\ref{fig:5}, indicating a continuous transformation from the bang-bang protocols corresponding to $\lambda=0$. (The protocols for ramping down $\Delta_3$ in this step are reflected about the center with a similar timescale.)  For finite $\lambda$, the sudden jumps are spread over finite time scales.

The overall protocol then looks very similar to the bang-bang protocol of Fig.~\ref{fig:1}(c) except each sudden jump is spread over a time window of length $\tau'\ll\tau^*$. We need to increase the total time of the operation by the sum of these ramp times to  to get a small final error. For example, in Fig. \ref{fig:5}, when we add 10\% to the time of step A, the protocol with a negligibly small ${\cal F}_A$ (for $\lambda=10^{-4}$) spreads the jump over a time interval of approximately $\tau^*_A/10$. 

We expect the smoothness of the protocol over timescales of order $1/E$, where $E$ is a large but finite energy gap to higher excitations [neglected in Hamiltonian Eq.~\eqref{eq:hamil0}], to prohibit leakage to these excitations. We may compare this timescale to that in a more realistic model. The gap may be associated with low energy Andreev bound states in some of the 
several junctions which are part of the circuits. It may also originate from finite Josephson ($E_J$) and charging energy ($E_C$) of the mesoscopic superconducting islands, with $E\sim \sqrt{E_CE_J}$.~\cite{aasen-milestones-2016}

\begin{figure}[]
	\includegraphics[width=7.5cm]{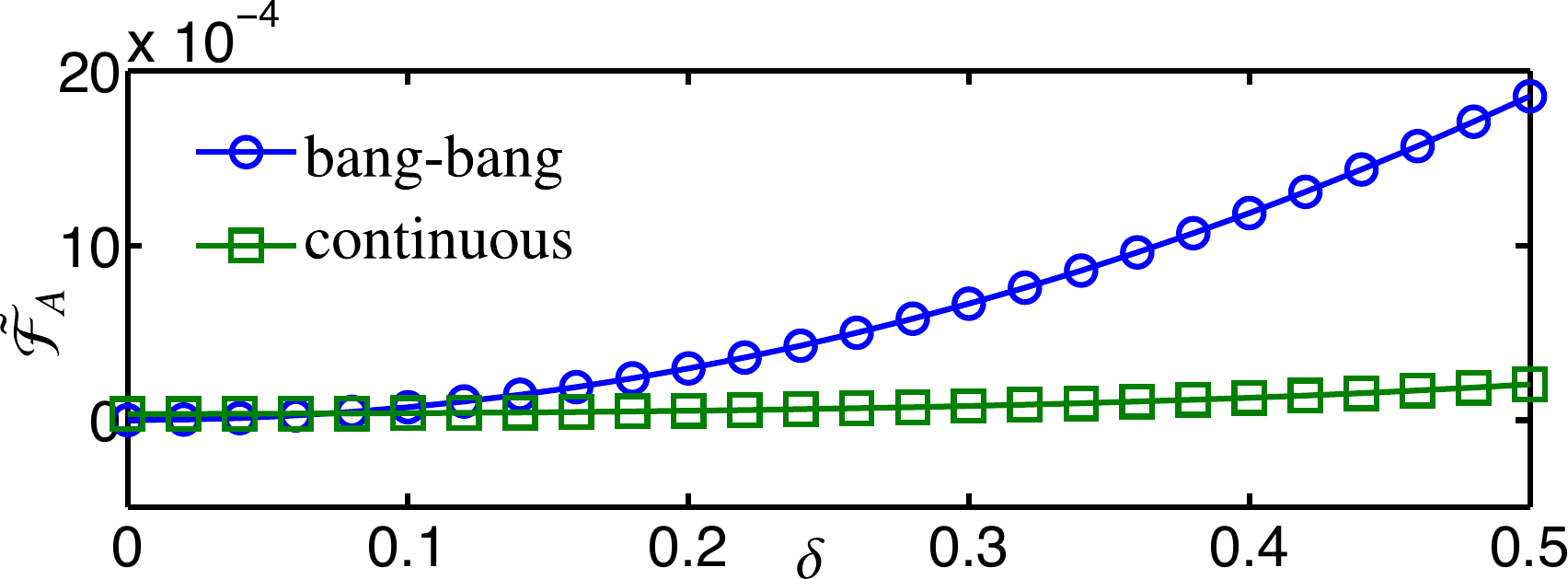}
	\caption{(Color online) The cost function $\tilde{\cal F}_A$ as a function of tunneling matrix element $\delta$ to a high-energy mode with $E=20$, for the bang-bag protocol with time $\tau^*_A$ and the continuous protocol of Fig.~\ref{fig:5} with time $t=1.1\tau^*_A$ and leakage-free cost function ${\cal F}_A=0.000035$. As expected, the continuous protocol exhibits significant advantages in the presence of leakage.
	\label{fig:6}}
\end{figure}

In practice, this energy scale is much larger than ${\cal D}_j$. We can use a simple toy model with one more mode to compare the smooth and bang-bang protocols and quantify the advantages of using the smooth protocols obtained above. Focusing on step A of the process with ${\cal H}=\Delta_1\sigma_y-\Delta_3\sigma_z$, we enlarge the two-dimensional Hilbert space to a three-dimensional space with 
\begin{equation}
\tilde {\cal H}=\left(\begin{array}{ccc}
-\Delta_3 & -i\Delta_1 & \delta\\ 
i\Delta_1 & \Delta_3 & \delta \\ 
\delta & \delta & E
\end{array} \right),
\end{equation}
where $\delta$ is a tunneling matrix element to the high energy mode. To account for leakage, we compute the cost function $\tilde{\cal F}_A$, using the $3\times 3$ $\tilde {\cal H}$ above and three-dimensional vectors for $|+z\rangle$ and $|-y\rangle$ with a vanishing third element [see Eq. \eqref{eq:F_A}]. Focusing on the protocol in Fig. \ref{fig:5} with $\lambda=10^{-4}$ and setting $E=20$, we computed $\tilde{\cal F}_A$ as a function of $\delta$. As seen in Fig. \ref{fig:6}, the results demonstrate the adiabatic suppression of leakage when using the smooth protocol. Note that for $\delta=0$, there cannot be any leakage and we obtain the previous cost functions (zero for the bang-bang protocol and 0.000035 for the smooth protocol).

\section{conclusions}
\label{sec:conc}
In summary, based on Pontryagin's theorem of optimal control, we proposed the optimal protocols for generating the same unitary operator as the one corresponding to fully adiabatic braiding of MZMs. While not providing full topological protection, our constrained optimal-control approach makes use of the nonlocal nature of the information stored in MZMs to make the system robust against a range of environmental perturbations. Through tailored diabatic pulse shapes, our scheme can significantly increase the speed of devices such as the top-transmon, without the need for any change to the experimental setup. Such fast accurate operations may defend the system against decoherence effects such as quasiparticle poisoning. The advantages of our method survive in the presence of white and $1/f$ noise and small calibration errors. The robustness can be further enhanced by making the pulses continuous without significantly sacrificing the performance of the device. Our proposed optimal diabatic gates can foster the development of high-performance quantum information processing with MZMs.

\acknowledgements{This work was supported by NSERC (M.F. and A.R.), CIfAR (M.F.), Max Planck-UBC Centre for Quantum Materials (M.F. and A.R.), and by the NSF CAREER award under Grant No. DMR-1350663, the U.S.-Israel Binational Science Foundation under Grant No. 2014345, as well as the College of Arts and Sciences at Indiana University (B.S.). We acknowledge support provided by WestGrid (www.westgrid.ca) and Compute Canada Calcul Canada (www.computecanada.ca).}

\bibliography{majorana.bib}

\begin{thebibliography}{46}%
\makeatletter
\providecommand \@ifxundefined [1]{%
 \@ifx{#1\undefined}
}%
\providecommand \@ifnum [1]{%
 \ifnum #1\expandafter \@firstoftwo
 \else \expandafter \@secondoftwo
 \fi
}%
\providecommand \@ifx [1]{%
 \ifx #1\expandafter \@firstoftwo
 \else \expandafter \@secondoftwo
 \fi
}%
\providecommand \natexlab [1]{#1}%
\providecommand \enquote  [1]{``#1''}%
\providecommand \bibnamefont  [1]{#1}%
\providecommand \bibfnamefont [1]{#1}%
\providecommand \citenamefont [1]{#1}%
\providecommand \href@noop [0]{\@secondoftwo}%
\providecommand \href [0]{\begingroup \@sanitize@url \@href}%
\providecommand \@href[1]{\@@startlink{#1}\@@href}%
\providecommand \@@href[1]{\endgroup#1\@@endlink}%
\providecommand \@sanitize@url [0]{\catcode `\\12\catcode `\$12\catcode
  `\&12\catcode `\#12\catcode `\^12\catcode `\_12\catcode `\%12\relax}%
\providecommand \@@startlink[1]{}%
\providecommand \@@endlink[0]{}%
\providecommand \url  [0]{\begingroup\@sanitize@url \@url }%
\providecommand \@url [1]{\endgroup\@href {#1}{\urlprefix }}%
\providecommand \urlprefix  [0]{URL }%
\providecommand \Eprint [0]{\href }%
\providecommand \doibase [0]{http://dx.doi.org/}%
\providecommand \selectlanguage [0]{\@gobble}%
\providecommand \bibinfo  [0]{\@secondoftwo}%
\providecommand \bibfield  [0]{\@secondoftwo}%
\providecommand \translation [1]{[#1]}%
\providecommand \BibitemOpen [0]{}%
\providecommand \bibitemStop [0]{}%
\providecommand \bibitemNoStop [0]{.\EOS\space}%
\providecommand \EOS [0]{\spacefactor3000\relax}%
\providecommand \BibitemShut  [1]{\csname bibitem#1\endcsname}%
\let\auto@bib@innerbib\@empty
%</preamble>
\bibitem [{\citenamefont {Kitaev}(2003)}]{kitaev_fault-tolerant_2003}%
  \BibitemOpen
  \bibfield  {author} {\bibinfo {author} {\bibfnamefont {A.}~\bibnamefont
  {Kitaev}},\ }\href {\doibase 10.1016/S0003-4916(02)00018-0} {\bibfield
  {journal} {\bibinfo  {journal} {Ann. Phys.}\ }\textbf {\bibinfo {volume}
  {303}},\ \bibinfo {pages} {2} (\bibinfo {year} {2003})}\BibitemShut {NoStop}%
\bibitem [{\citenamefont {Nayak}\ \emph {et~al.}(2008)\citenamefont {Nayak},
  \citenamefont {Simon}, \citenamefont {Stern}, \citenamefont {Freedman},\ and\
  \citenamefont {Das~Sarma}}]{nayak_non-abelian_2008}%
  \BibitemOpen
  \bibfield  {author} {\bibinfo {author} {\bibfnamefont {C.}~\bibnamefont
  {Nayak}}, \bibinfo {author} {\bibfnamefont {S.~H.}\ \bibnamefont {Simon}},
  \bibinfo {author} {\bibfnamefont {A.}~\bibnamefont {Stern}}, \bibinfo
  {author} {\bibfnamefont {M.}~\bibnamefont {Freedman}}, \ and\ \bibinfo
  {author} {\bibfnamefont {S.}~\bibnamefont {Das~Sarma}},\ }\href {\doibase
  10.1103/RevModPhys.80.1083} {\bibfield  {journal} {\bibinfo  {journal} {Rev.
  Mod. Phys.}\ }\textbf {\bibinfo {volume} {80}},\ \bibinfo {pages} {1083}
  (\bibinfo {year} {2008})}\BibitemShut {NoStop}%
\bibitem [{\citenamefont {Oreg}\ \emph {et~al.}(2010)\citenamefont {Oreg},
  \citenamefont {Refael},\ and\ \citenamefont {von Oppen}}]{oreg_helical_2010}%
  \BibitemOpen
  \bibfield  {author} {\bibinfo {author} {\bibfnamefont {Y.}~\bibnamefont
  {Oreg}}, \bibinfo {author} {\bibfnamefont {G.}~\bibnamefont {Refael}}, \ and\
  \bibinfo {author} {\bibfnamefont {F.}~\bibnamefont {von Oppen}},\ }\href
  {\doibase 10.1103/PhysRevLett.105.177002} {\bibfield  {journal} {\bibinfo
  {journal} {Phys. Rev. Lett.}\ }\textbf {\bibinfo {volume} {105}},\ \bibinfo
  {pages} {177002} (\bibinfo {year} {2010})}\BibitemShut {NoStop}%
\bibitem [{\citenamefont {Lutchyn}\ \emph {et~al.}(2010)\citenamefont
  {Lutchyn}, \citenamefont {Sau},\ and\ \citenamefont
  {Das~Sarma}}]{lutchyn_majorana_2010}%
  \BibitemOpen
  \bibfield  {author} {\bibinfo {author} {\bibfnamefont {R.~M.}\ \bibnamefont
  {Lutchyn}}, \bibinfo {author} {\bibfnamefont {J.~D.}\ \bibnamefont {Sau}}, \
  and\ \bibinfo {author} {\bibfnamefont {S.}~\bibnamefont {Das~Sarma}},\ }\href
  {\doibase 10.1103/PhysRevLett.105.077001} {\bibfield  {journal} {\bibinfo
  {journal} {Phys. Rev. Lett.}\ }\textbf {\bibinfo {volume} {105}},\ \bibinfo
  {pages} {077001} (\bibinfo {year} {2010})}\BibitemShut {NoStop}%
\bibitem [{\citenamefont {Alicea}(2012)}]{alicea_new_2012}%
  \BibitemOpen
  \bibfield  {author} {\bibinfo {author} {\bibfnamefont {J.}~\bibnamefont
  {Alicea}},\ }\href
  {http://iopscience.iop.org/article/10.1088/0034-4885/75/7/076501/meta}
  {\bibfield  {journal} {\bibinfo  {journal} {Rep. Prog. Phys.}\ }\textbf
  {\bibinfo {volume} {75}},\ \bibinfo {pages} {076501} (\bibinfo {year}
  {2012})}\BibitemShut {NoStop}%
\bibitem [{\citenamefont {Beenakker}(2013)}]{beenakker_2012}%
  \BibitemOpen
  \bibfield  {author} {\bibinfo {author} {\bibfnamefont {C.}~\bibnamefont
  {Beenakker}},\ }\href
  {http://www.annualreviews.org/doi/abs/10.1146/annurev-conmatphys-030212-184337}
  {\bibfield  {journal} {\bibinfo  {journal} {Annu. Rev. Con. Mat. Phys.}\
  }\textbf {\bibinfo {volume} {4}},\ \bibinfo {pages} {113} (\bibinfo {year}
  {2013})}\BibitemShut {NoStop}%
\bibitem [{\citenamefont {Elliott}\ and\ \citenamefont
  {Franz}(2015)}]{elliott_majorana_2015}%
  \BibitemOpen
  \bibfield  {author} {\bibinfo {author} {\bibfnamefont {S.~R.}\ \bibnamefont
  {Elliott}}\ and\ \bibinfo {author} {\bibfnamefont {M.}~\bibnamefont
  {Franz}},\ }\href {\doibase 10.1103/RevModPhys.87.137} {\bibfield  {journal}
  {\bibinfo  {journal} {Rev. Mod. Phys.}\ }\textbf {\bibinfo {volume} {87}},\
  \bibinfo {pages} {137} (\bibinfo {year} {2015})}\BibitemShut {NoStop}%
\bibitem [{\citenamefont {Mourik}\ \emph {et~al.}(2012)\citenamefont {Mourik},
  \citenamefont {Zuo}, \citenamefont {Frolov}, \citenamefont {Plissard},
  \citenamefont {Bakkers},\ and\ \citenamefont
  {Kouwenhoven}}]{mourik_signatures_2012}%
  \BibitemOpen
  \bibfield  {author} {\bibinfo {author} {\bibfnamefont {V.}~\bibnamefont
  {Mourik}}, \bibinfo {author} {\bibfnamefont {K.}~\bibnamefont {Zuo}},
  \bibinfo {author} {\bibfnamefont {S.~M.}\ \bibnamefont {Frolov}}, \bibinfo
  {author} {\bibfnamefont {S.~R.}\ \bibnamefont {Plissard}}, \bibinfo {author}
  {\bibfnamefont {E.~P. a.~M.}\ \bibnamefont {Bakkers}}, \ and\ \bibinfo
  {author} {\bibfnamefont {L.~P.}\ \bibnamefont {Kouwenhoven}},\ }\href
  {\doibase 10.1126/science.1222360} {\bibfield  {journal} {\bibinfo  {journal}
  {Science}\ }\textbf {\bibinfo {volume} {336}},\ \bibinfo {pages} {1003}
  (\bibinfo {year} {2012})}\BibitemShut {NoStop}%
\bibitem [{\citenamefont {Das}\ \emph {et~al.}(2012)\citenamefont {Das},
  \citenamefont {Ronen}, \citenamefont {Most}, \citenamefont {Oreg},
  \citenamefont {Heiblum},\ and\ \citenamefont
  {Shtrikman}}]{das_zero-bias_2012}%
  \BibitemOpen
  \bibfield  {author} {\bibinfo {author} {\bibfnamefont {A.}~\bibnamefont
  {Das}}, \bibinfo {author} {\bibfnamefont {Y.}~\bibnamefont {Ronen}}, \bibinfo
  {author} {\bibfnamefont {Y.}~\bibnamefont {Most}}, \bibinfo {author}
  {\bibfnamefont {Y.}~\bibnamefont {Oreg}}, \bibinfo {author} {\bibfnamefont
  {M.}~\bibnamefont {Heiblum}}, \ and\ \bibinfo {author} {\bibfnamefont
  {H.}~\bibnamefont {Shtrikman}},\ }\href {\doibase 10.1038/nphys2479}
  {\bibfield  {journal} {\bibinfo  {journal} {Nat. Phys.}\ }\textbf {\bibinfo
  {volume} {8}},\ \bibinfo {pages} {887} (\bibinfo {year} {2012})}\BibitemShut
  {NoStop}%
\bibitem [{\citenamefont {Churchill}\ \emph {et~al.}(2013)\citenamefont
  {Churchill}, \citenamefont {Fatemi}, \citenamefont {Grove-Rasmussen},
  \citenamefont {Deng}, \citenamefont {Caroff}, \citenamefont {Xu},\ and\
  \citenamefont {Marcus}}]{churchill_superconductor_2013}%
  \BibitemOpen
  \bibfield  {author} {\bibinfo {author} {\bibfnamefont {H.~O.~H.}\
  \bibnamefont {Churchill}}, \bibinfo {author} {\bibfnamefont {V.}~\bibnamefont
  {Fatemi}}, \bibinfo {author} {\bibfnamefont {K.}~\bibnamefont
  {Grove-Rasmussen}}, \bibinfo {author} {\bibfnamefont {M.~T.}\ \bibnamefont
  {Deng}}, \bibinfo {author} {\bibfnamefont {P.}~\bibnamefont {Caroff}},
  \bibinfo {author} {\bibfnamefont {H.~Q.}\ \bibnamefont {Xu}}, \ and\ \bibinfo
  {author} {\bibfnamefont {C.~M.}\ \bibnamefont {Marcus}},\ }\href {\doibase
  10.1103/PhysRevB.87.241401} {\bibfield  {journal} {\bibinfo  {journal} {Phys.
  Rev. B}\ }\textbf {\bibinfo {volume} {87}},\ \bibinfo {pages} {241401}
  (\bibinfo {year} {2013})}\BibitemShut {NoStop}%
\bibitem [{\citenamefont {Rokhinson}\ \emph {et~al.}(2012)\citenamefont
  {Rokhinson}, \citenamefont {Liu},\ and\ \citenamefont
  {Furdyna}}]{rokhinson_fractional_2012}%
  \BibitemOpen
  \bibfield  {author} {\bibinfo {author} {\bibfnamefont {L.~P.}\ \bibnamefont
  {Rokhinson}}, \bibinfo {author} {\bibfnamefont {X.}~\bibnamefont {Liu}}, \
  and\ \bibinfo {author} {\bibfnamefont {J.~K.}\ \bibnamefont {Furdyna}},\
  }\href {\doibase 10.1038/nphys2429} {\bibfield  {journal} {\bibinfo
  {journal} {Nat Phys}\ }\textbf {\bibinfo {volume} {8}},\ \bibinfo {pages}
  {795} (\bibinfo {year} {2012})}\BibitemShut {NoStop}%
\bibitem [{\citenamefont {Deng}\ \emph {et~al.}(2012)\citenamefont {Deng},
  \citenamefont {Yu}, \citenamefont {Huang}, \citenamefont {Larsson},
  \citenamefont {Caroff},\ and\ \citenamefont {Xu}}]{deng_anomalous_2012}%
  \BibitemOpen
  \bibfield  {author} {\bibinfo {author} {\bibfnamefont {M.~T.}\ \bibnamefont
  {Deng}}, \bibinfo {author} {\bibfnamefont {C.~L.}\ \bibnamefont {Yu}},
  \bibinfo {author} {\bibfnamefont {G.~Y.}\ \bibnamefont {Huang}}, \bibinfo
  {author} {\bibfnamefont {M.}~\bibnamefont {Larsson}}, \bibinfo {author}
  {\bibfnamefont {P.}~\bibnamefont {Caroff}}, \ and\ \bibinfo {author}
  {\bibfnamefont {H.~Q.}\ \bibnamefont {Xu}},\ }\href {\doibase
  10.1021/nl303758w} {\bibfield  {journal} {\bibinfo  {journal} {Nano Lett.}\
  }\textbf {\bibinfo {volume} {12}},\ \bibinfo {pages} {6414} (\bibinfo {year}
  {2012})}\BibitemShut {NoStop}%
\bibitem [{\citenamefont {Finck}\ \emph {et~al.}(2013)\citenamefont {Finck},
  \citenamefont {Van~Harlingen}, \citenamefont {Mohseni}, \citenamefont
  {Jung},\ and\ \citenamefont {Li}}]{finck_anomalous_2013}%
  \BibitemOpen
  \bibfield  {author} {\bibinfo {author} {\bibfnamefont {A.~D.~K.}\
  \bibnamefont {Finck}}, \bibinfo {author} {\bibfnamefont {D.~J.}\ \bibnamefont
  {Van~Harlingen}}, \bibinfo {author} {\bibfnamefont {P.~K.}\ \bibnamefont
  {Mohseni}}, \bibinfo {author} {\bibfnamefont {K.}~\bibnamefont {Jung}}, \
  and\ \bibinfo {author} {\bibfnamefont {X.}~\bibnamefont {Li}},\ }\href
  {\doibase 10.1103/PhysRevLett.110.126406} {\bibfield  {journal} {\bibinfo
  {journal} {Phys. Rev. Lett.}\ }\textbf {\bibinfo {volume} {110}},\ \bibinfo
  {pages} {126406} (\bibinfo {year} {2013})}\BibitemShut {NoStop}%
\bibitem [{\citenamefont {Nadj-Perge}\ \emph {et~al.}(2014)\citenamefont
  {Nadj-Perge}, \citenamefont {Drozdov}, \citenamefont {Li}, \citenamefont
  {Chen}, \citenamefont {Jeon}, \citenamefont {Seo}, \citenamefont {MacDonald},
  \citenamefont {Bernevig},\ and\ \citenamefont
  {Yazdani}}]{nadj-perge_observation_2014}%
  \BibitemOpen
  \bibfield  {author} {\bibinfo {author} {\bibfnamefont {S.}~\bibnamefont
  {Nadj-Perge}}, \bibinfo {author} {\bibfnamefont {I.~K.}\ \bibnamefont
  {Drozdov}}, \bibinfo {author} {\bibfnamefont {J.}~\bibnamefont {Li}},
  \bibinfo {author} {\bibfnamefont {H.}~\bibnamefont {Chen}}, \bibinfo {author}
  {\bibfnamefont {S.}~\bibnamefont {Jeon}}, \bibinfo {author} {\bibfnamefont
  {J.}~\bibnamefont {Seo}}, \bibinfo {author} {\bibfnamefont {A.~H.}\
  \bibnamefont {MacDonald}}, \bibinfo {author} {\bibfnamefont {B.~A.}\
  \bibnamefont {Bernevig}}, \ and\ \bibinfo {author} {\bibfnamefont
  {A.}~\bibnamefont {Yazdani}},\ }\href {\doibase 10.1126/science.1259327}
  {\bibfield  {journal} {\bibinfo  {journal} {Science}\ }\textbf {\bibinfo
  {volume} {346}},\ \bibinfo {pages} {602} (\bibinfo {year}
  {2014})}\BibitemShut {NoStop}%
\bibitem [{\citenamefont {Aasen}\ \emph {et~al.}()\citenamefont {Aasen},
  \citenamefont {Hell}, \citenamefont {Mishmash}, \citenamefont {Higginbotham},
  \citenamefont {Danon}, \citenamefont {Leijnse}, \citenamefont {Jespersen},
  \citenamefont {Folk}, \citenamefont {Marcus}, \citenamefont {Flensberg},\
  and\ \citenamefont {Alicea}}]{aasen-milestones-2016}%
  \BibitemOpen
  \bibfield  {author} {\bibinfo {author} {\bibfnamefont {D.}~\bibnamefont
  {Aasen}}, \bibinfo {author} {\bibfnamefont {M.}~\bibnamefont {Hell}},
  \bibinfo {author} {\bibfnamefont {R.~V.}\ \bibnamefont {Mishmash}}, \bibinfo
  {author} {\bibfnamefont {A.}~\bibnamefont {Higginbotham}}, \bibinfo {author}
  {\bibfnamefont {J.}~\bibnamefont {Danon}}, \bibinfo {author} {\bibfnamefont
  {M.}~\bibnamefont {Leijnse}}, \bibinfo {author} {\bibfnamefont {T.~S.}\
  \bibnamefont {Jespersen}}, \bibinfo {author} {\bibfnamefont {J.~A.}\
  \bibnamefont {Folk}}, \bibinfo {author} {\bibfnamefont {C.~M.}\ \bibnamefont
  {Marcus}}, \bibinfo {author} {\bibfnamefont {K.}~\bibnamefont {Flensberg}}, \
  and\ \bibinfo {author} {\bibfnamefont {J.}~\bibnamefont {Alicea}},\
  }\href@noop {} {}\Eprint {http://arxiv.org/abs/1511.05153} {arXiv:1511.05153}
  \BibitemShut {NoStop}%
\bibitem [{\citenamefont {Cheng}\ \emph {et~al.}(2011)\citenamefont {Cheng},
  \citenamefont {Galitski},\ and\ \citenamefont
  {Das~Sarma}}]{cheng_nonadiabatic_2011}%
  \BibitemOpen
  \bibfield  {author} {\bibinfo {author} {\bibfnamefont {M.}~\bibnamefont
  {Cheng}}, \bibinfo {author} {\bibfnamefont {V.}~\bibnamefont {Galitski}}, \
  and\ \bibinfo {author} {\bibfnamefont {S.}~\bibnamefont {Das~Sarma}},\ }\href
  {\doibase 10.1103/PhysRevB.84.104529} {\bibfield  {journal} {\bibinfo
  {journal} {Phys. Rev. B}\ }\textbf {\bibinfo {volume} {84}},\ \bibinfo
  {pages} {104529} (\bibinfo {year} {2011})}\BibitemShut {NoStop}%
\bibitem [{\citenamefont {Karzig}\ \emph {et~al.}(2013)\citenamefont {Karzig},
  \citenamefont {Refael},\ and\ \citenamefont {von
  Oppen}}]{karzig_boosting_2013}%
  \BibitemOpen
  \bibfield  {author} {\bibinfo {author} {\bibfnamefont {T.}~\bibnamefont
  {Karzig}}, \bibinfo {author} {\bibfnamefont {G.}~\bibnamefont {Refael}}, \
  and\ \bibinfo {author} {\bibfnamefont {F.}~\bibnamefont {von Oppen}},\ }\href
  {\doibase 10.1103/PhysRevX.3.041017} {\bibfield  {journal} {\bibinfo
  {journal} {Phys. Rev. X}\ }\textbf {\bibinfo {volume} {3}},\ \bibinfo {pages}
  {041017} (\bibinfo {year} {2013})}\BibitemShut {NoStop}%
\bibitem [{\citenamefont {Amorim}\ \emph {et~al.}(2015)\citenamefont {Amorim},
  \citenamefont {Ebihara}, \citenamefont {Yamakage}, \citenamefont {Tanaka},\
  and\ \citenamefont {Sato}}]{amorim_majorana_2014}%
  \BibitemOpen
  \bibfield  {author} {\bibinfo {author} {\bibfnamefont {C.~S.}\ \bibnamefont
  {Amorim}}, \bibinfo {author} {\bibfnamefont {K.}~\bibnamefont {Ebihara}},
  \bibinfo {author} {\bibfnamefont {A.}~\bibnamefont {Yamakage}}, \bibinfo
  {author} {\bibfnamefont {Y.}~\bibnamefont {Tanaka}}, \ and\ \bibinfo {author}
  {\bibfnamefont {M.}~\bibnamefont {Sato}},\ }\href {\doibase
  10.1103/PhysRevB.91.174305} {\bibfield  {journal} {\bibinfo  {journal} {Phys.
  Rev. B}\ }\textbf {\bibinfo {volume} {91}},\ \bibinfo {pages} {174305}
  (\bibinfo {year} {2015})}\BibitemShut {NoStop}%
\bibitem [{\citenamefont {Rainis}\ and\ \citenamefont
  {Loss}(2012)}]{rainis_majorana_2012}%
  \BibitemOpen
  \bibfield  {author} {\bibinfo {author} {\bibfnamefont {D.}~\bibnamefont
  {Rainis}}\ and\ \bibinfo {author} {\bibfnamefont {D.}~\bibnamefont {Loss}},\
  }\href {\doibase 10.1103/PhysRevB.85.174533} {\bibfield  {journal} {\bibinfo
  {journal} {Phys. Rev. B}\ }\textbf {\bibinfo {volume} {85}},\ \bibinfo
  {pages} {174533} (\bibinfo {year} {2012})}\BibitemShut {NoStop}%
\bibitem [{\citenamefont {van Heck}\ \emph {et~al.}(2012)\citenamefont {van
  Heck}, \citenamefont {Akhmerov}, \citenamefont {Hassler}, \citenamefont
  {Burrello},\ and\ \citenamefont {Beenakker}}]{vanheck_coulomb_2012}%
  \BibitemOpen
  \bibfield  {author} {\bibinfo {author} {\bibfnamefont {B.}~\bibnamefont {van
  Heck}}, \bibinfo {author} {\bibfnamefont {A.~R.}\ \bibnamefont {Akhmerov}},
  \bibinfo {author} {\bibfnamefont {F.}~\bibnamefont {Hassler}}, \bibinfo
  {author} {\bibfnamefont {M.}~\bibnamefont {Burrello}}, \ and\ \bibinfo
  {author} {\bibfnamefont {C.~W.~J.}\ \bibnamefont {Beenakker}},\ }\href
  {http://stacks.iop.org/1367-2630/14/i=3/a=035019} {\bibfield  {journal}
  {\bibinfo  {journal} {New J. Phys.}\ }\textbf {\bibinfo {volume} {14}},\
  \bibinfo {pages} {035019} (\bibinfo {year} {2012})}\BibitemShut {NoStop}%
\bibitem [{\citenamefont {Goldstein}\ and\ \citenamefont
  {Chamon}(2011)}]{goldstein_decay_2011}%
  \BibitemOpen
  \bibfield  {author} {\bibinfo {author} {\bibfnamefont {G.}~\bibnamefont
  {Goldstein}}\ and\ \bibinfo {author} {\bibfnamefont {C.}~\bibnamefont
  {Chamon}},\ }\href {\doibase 10.1103/PhysRevB.84.205109} {\bibfield
  {journal} {\bibinfo  {journal} {Phys. Rev. B}\ }\textbf {\bibinfo {volume}
  {84}},\ \bibinfo {pages} {205109} (\bibinfo {year} {2011})}\BibitemShut
  {NoStop}%
\bibitem [{\citenamefont {Schmidt}\ \emph {et~al.}(2012)\citenamefont
  {Schmidt}, \citenamefont {Rainis},\ and\ \citenamefont
  {Loss}}]{schmidt_decoherence_2012}%
  \BibitemOpen
  \bibfield  {author} {\bibinfo {author} {\bibfnamefont {M.~J.}\ \bibnamefont
  {Schmidt}}, \bibinfo {author} {\bibfnamefont {D.}~\bibnamefont {Rainis}}, \
  and\ \bibinfo {author} {\bibfnamefont {D.}~\bibnamefont {Loss}},\ }\href
  {\doibase 10.1103/PhysRevB.86.085414} {\bibfield  {journal} {\bibinfo
  {journal} {Phys. Rev. B}\ }\textbf {\bibinfo {volume} {86}},\ \bibinfo
  {pages} {085414} (\bibinfo {year} {2012})}\BibitemShut {NoStop}%
\bibitem [{\citenamefont {Berry}(2009)}]{berry_transitionless_2009}%
  \BibitemOpen
  \bibfield  {author} {\bibinfo {author} {\bibfnamefont {M.~V.}\ \bibnamefont
  {Berry}},\ }\href {http://stacks.iop.org/1751-8121/42/i=36/a=365303}
  {\bibfield  {journal} {\bibinfo  {journal} {J. Phys. A: Math. Theor.}\
  }\textbf {\bibinfo {volume} {42}},\ \bibinfo {pages} {365303} (\bibinfo
  {year} {2009})}\BibitemShut {NoStop}%
\bibitem [{\citenamefont {del Campo}(2013)}]{delcampo_shortcuts_2013}%
  \BibitemOpen
  \bibfield  {author} {\bibinfo {author} {\bibfnamefont {A.}~\bibnamefont {del
  Campo}},\ }\href {\doibase 10.1103/PhysRevLett.111.100502} {\bibfield
  {journal} {\bibinfo  {journal} {Phys. Rev. Lett.}\ }\textbf {\bibinfo
  {volume} {111}},\ \bibinfo {pages} {100502} (\bibinfo {year}
  {2013})}\BibitemShut {NoStop}%
\bibitem [{\citenamefont {Karzig}\ \emph
  {et~al.}(2015{\natexlab{a}})\citenamefont {Karzig}, \citenamefont {Pientka},
  \citenamefont {Refael},\ and\ \citenamefont {von
  Oppen}}]{karzig_shortcut_2015}%
  \BibitemOpen
  \bibfield  {author} {\bibinfo {author} {\bibfnamefont {T.}~\bibnamefont
  {Karzig}}, \bibinfo {author} {\bibfnamefont {F.}~\bibnamefont {Pientka}},
  \bibinfo {author} {\bibfnamefont {G.}~\bibnamefont {Refael}}, \ and\ \bibinfo
  {author} {\bibfnamefont {F.}~\bibnamefont {von Oppen}},\ }\href {\doibase
  10.1103/PhysRevB.91.201102} {\bibfield  {journal} {\bibinfo  {journal} {Phys.
  Rev. B}\ }\textbf {\bibinfo {volume} {91}},\ \bibinfo {pages} {201102}
  (\bibinfo {year} {2015}{\natexlab{a}})}\BibitemShut {NoStop}%
\bibitem [{\citenamefont {Zhang}\ \emph {et~al.}(2015)\citenamefont {Zhang},
  \citenamefont {Kyaw}, \citenamefont {Tong}, \citenamefont {Sj\"oqvist},\ and\
  \citenamefont {Kwek}}]{zhang_shortcut_2015}%
  \BibitemOpen
  \bibfield  {author} {\bibinfo {author} {\bibfnamefont {J.}~\bibnamefont
  {Zhang}}, \bibinfo {author} {\bibfnamefont {T.~H.}\ \bibnamefont {Kyaw}},
  \bibinfo {author} {\bibfnamefont {D.~M.}\ \bibnamefont {Tong}}, \bibinfo
  {author} {\bibfnamefont {E.}~\bibnamefont {Sj\"oqvist}}, \ and\ \bibinfo
  {author} {\bibfnamefont {L.-C.}\ \bibnamefont {Kwek}},\ }\href
  {http://www.nature.com/articles/srep18414} {\bibfield  {journal} {\bibinfo
  {journal} {Sci. Rep.}\ }\textbf {\bibinfo {volume} {5}},\ \bibinfo {pages}
  {18414} (\bibinfo {year} {2015})}\BibitemShut {NoStop}%
\bibitem [{\citenamefont {Knapp}\ \emph {et~al.}(2016)\citenamefont {Knapp},
  \citenamefont {Zaletel}, \citenamefont {Liu}, \citenamefont {Cheng},
  \citenamefont {Bonderson},\ and\ \citenamefont
  {Nayak}}]{knap_quick_braid_2016}%
  \BibitemOpen
  \bibfield  {author} {\bibinfo {author} {\bibfnamefont {C.}~\bibnamefont
  {Knapp}}, \bibinfo {author} {\bibfnamefont {M.}~\bibnamefont {Zaletel}},
  \bibinfo {author} {\bibfnamefont {D.~E.}\ \bibnamefont {Liu}}, \bibinfo
  {author} {\bibfnamefont {M.}~\bibnamefont {Cheng}}, \bibinfo {author}
  {\bibfnamefont {P.}~\bibnamefont {Bonderson}}, \ and\ \bibinfo {author}
  {\bibfnamefont {C.}~\bibnamefont {Nayak}},\ }\href {\doibase
  10.1103/PhysRevX.6.041003} {\bibfield  {journal} {\bibinfo  {journal} {Phys.
  Rev. X}\ }\textbf {\bibinfo {volume} {6}},\ \bibinfo {pages} {041003}
  (\bibinfo {year} {2016})}\BibitemShut {NoStop}%
\bibitem [{\citenamefont {Peirce}\ \emph {et~al.}(1988)\citenamefont {Peirce},
  \citenamefont {Dahleh},\ and\ \citenamefont {Rabitz}}]{peirce_optimal_1988}%
  \BibitemOpen
  \bibfield  {author} {\bibinfo {author} {\bibfnamefont {A.~P.}\ \bibnamefont
  {Peirce}}, \bibinfo {author} {\bibfnamefont {M.~A.}\ \bibnamefont {Dahleh}},
  \ and\ \bibinfo {author} {\bibfnamefont {H.}~\bibnamefont {Rabitz}},\ }\href
  {\doibase 10.1103/PhysRevA.37.4950} {\bibfield  {journal} {\bibinfo
  {journal} {Phys. Rev. A}\ }\textbf {\bibinfo {volume} {37}},\ \bibinfo
  {pages} {4950} (\bibinfo {year} {1988})}\BibitemShut {NoStop}%
\bibitem [{\citenamefont {Palao}\ and\ \citenamefont
  {Kosloff}(2002)}]{palao_quantum_2002}%
  \BibitemOpen
  \bibfield  {author} {\bibinfo {author} {\bibfnamefont {J.~P.}\ \bibnamefont
  {Palao}}\ and\ \bibinfo {author} {\bibfnamefont {R.}~\bibnamefont
  {Kosloff}},\ }\href {\doibase 10.1103/PhysRevLett.89.188301} {\bibfield
  {journal} {\bibinfo  {journal} {Phys. Rev. Lett.}\ }\textbf {\bibinfo
  {volume} {89}},\ \bibinfo {pages} {188301} (\bibinfo {year}
  {2002})}\BibitemShut {NoStop}%
\bibitem [{\citenamefont {Kr\'al}\ \emph {et~al.}(2007)\citenamefont {Kr\'al},
  \citenamefont {Thanopulos},\ and\ \citenamefont
  {Shapiro}}]{kral_coherently_2007}%
  \BibitemOpen
  \bibfield  {author} {\bibinfo {author} {\bibfnamefont {P.}~\bibnamefont
  {Kr\'al}}, \bibinfo {author} {\bibfnamefont {I.}~\bibnamefont {Thanopulos}},
  \ and\ \bibinfo {author} {\bibfnamefont {M.}~\bibnamefont {Shapiro}},\ }\href
  {\doibase 10.1103/RevModPhys.79.53} {\bibfield  {journal} {\bibinfo
  {journal} {Rev. Mod. Phys.}\ }\textbf {\bibinfo {volume} {79}},\ \bibinfo
  {pages} {53} (\bibinfo {year} {2007})}\BibitemShut {NoStop}%
\bibitem [{\citenamefont {Caneva}\ \emph {et~al.}(2009)\citenamefont {Caneva},
  \citenamefont {Murphy}, \citenamefont {Calarco}, \citenamefont {Fazio},
  \citenamefont {Montangero}, \citenamefont {Giovannetti},\ and\ \citenamefont
  {Santoro}}]{caneva_optila_2009}%
  \BibitemOpen
  \bibfield  {author} {\bibinfo {author} {\bibfnamefont {T.}~\bibnamefont
  {Caneva}}, \bibinfo {author} {\bibfnamefont {M.}~\bibnamefont {Murphy}},
  \bibinfo {author} {\bibfnamefont {T.}~\bibnamefont {Calarco}}, \bibinfo
  {author} {\bibfnamefont {R.}~\bibnamefont {Fazio}}, \bibinfo {author}
  {\bibfnamefont {S.}~\bibnamefont {Montangero}}, \bibinfo {author}
  {\bibfnamefont {V.}~\bibnamefont {Giovannetti}}, \ and\ \bibinfo {author}
  {\bibfnamefont {G.~E.}\ \bibnamefont {Santoro}},\ }\href {\doibase
  10.1103/PhysRevLett.103.240501} {\bibfield  {journal} {\bibinfo  {journal}
  {Phys. Rev. Lett.}\ }\textbf {\bibinfo {volume} {103}},\ \bibinfo {pages}
  {240501} (\bibinfo {year} {2009})}\BibitemShut {NoStop}%
\bibitem [{\citenamefont {Doria}\ \emph {et~al.}(2011)\citenamefont {Doria},
  \citenamefont {Calarco},\ and\ \citenamefont
  {Montangero}}]{doria_optimal_2011}%
  \BibitemOpen
  \bibfield  {author} {\bibinfo {author} {\bibfnamefont {P.}~\bibnamefont
  {Doria}}, \bibinfo {author} {\bibfnamefont {T.}~\bibnamefont {Calarco}}, \
  and\ \bibinfo {author} {\bibfnamefont {S.}~\bibnamefont {Montangero}},\
  }\href {\doibase 10.1103/PhysRevLett.106.190501} {\bibfield  {journal}
  {\bibinfo  {journal} {Phys. Rev. Lett.}\ }\textbf {\bibinfo {volume} {106}},\
  \bibinfo {pages} {190501} (\bibinfo {year} {2011})}\BibitemShut {NoStop}%
\bibitem [{\citenamefont {Rahmani}\ and\ \citenamefont
  {Chamon}(2011)}]{rahmani_optimal_2011}%
  \BibitemOpen
  \bibfield  {author} {\bibinfo {author} {\bibfnamefont {A.}~\bibnamefont
  {Rahmani}}\ and\ \bibinfo {author} {\bibfnamefont {C.}~\bibnamefont
  {Chamon}},\ }\href {\doibase 10.1103/PhysRevLett.107.016402} {\bibfield
  {journal} {\bibinfo  {journal} {Phys. Rev. Lett.}\ }\textbf {\bibinfo
  {volume} {107}},\ \bibinfo {pages} {016402} (\bibinfo {year}
  {2011})}\BibitemShut {NoStop}%
\bibitem [{\citenamefont {Rahmani}(2013)}]{rahmani_quantum_2013}%
  \BibitemOpen
  \bibfield  {author} {\bibinfo {author} {\bibfnamefont {A.}~\bibnamefont
  {Rahmani}},\ }\href {\doibase 10.1142/S0217984913300196} {\bibfield
  {journal} {\bibinfo  {journal} {Mod. Phys. Lett. B}\ }\textbf {\bibinfo
  {volume} {27}},\ \bibinfo {pages} {1330019} (\bibinfo {year}
  {2013})}\BibitemShut {NoStop}%
\bibitem [{\citenamefont {Karzig}\ \emph
  {et~al.}(2015{\natexlab{b}})\citenamefont {Karzig}, \citenamefont {Rahmani},
  \citenamefont {von Oppen},\ and\ \citenamefont
  {Refael}}]{karzig_optimal_2015}%
  \BibitemOpen
  \bibfield  {author} {\bibinfo {author} {\bibfnamefont {T.}~\bibnamefont
  {Karzig}}, \bibinfo {author} {\bibfnamefont {A.}~\bibnamefont {Rahmani}},
  \bibinfo {author} {\bibfnamefont {F.}~\bibnamefont {von Oppen}}, \ and\
  \bibinfo {author} {\bibfnamefont {G.}~\bibnamefont {Refael}},\ }\href
  {\doibase 10.1103/PhysRevB.91.201404} {\bibfield  {journal} {\bibinfo
  {journal} {Phys. Rev. B}\ }\textbf {\bibinfo {volume} {91}},\ \bibinfo
  {pages} {201404} (\bibinfo {year} {2015}{\natexlab{b}})}\BibitemShut
  {NoStop}%
\bibitem [{\citenamefont {Hassler}\ \emph {et~al.}(2011)\citenamefont
  {Hassler}, \citenamefont {Akhmerov},\ and\ \citenamefont
  {Beenakker}}]{hassler_toptransmon_2011}%
  \BibitemOpen
  \bibfield  {author} {\bibinfo {author} {\bibfnamefont {F.}~\bibnamefont
  {Hassler}}, \bibinfo {author} {\bibfnamefont {A.~R.}\ \bibnamefont
  {Akhmerov}}, \ and\ \bibinfo {author} {\bibfnamefont {C.~W.~J.}\ \bibnamefont
  {Beenakker}},\ }\href {http://stacks.iop.org/1367-2630/13/i=9/a=095004}
  {\bibfield  {journal} {\bibinfo  {journal} {New J. Phys.}\ }\textbf {\bibinfo
  {volume} {13}},\ \bibinfo {pages} {095004} (\bibinfo {year}
  {2011})}\BibitemShut {NoStop}%
\bibitem [{\citenamefont {Hyart}\ \emph {et~al.}(2013)\citenamefont {Hyart},
  \citenamefont {van Heck}, \citenamefont {Fulga}, \citenamefont {Burrello},
  \citenamefont {Akhmerov},\ and\ \citenamefont {Beenakker}}]{hyart_flux_2013}%
  \BibitemOpen
  \bibfield  {author} {\bibinfo {author} {\bibfnamefont {T.}~\bibnamefont
  {Hyart}}, \bibinfo {author} {\bibfnamefont {B.}~\bibnamefont {van Heck}},
  \bibinfo {author} {\bibfnamefont {I.~C.}\ \bibnamefont {Fulga}}, \bibinfo
  {author} {\bibfnamefont {M.}~\bibnamefont {Burrello}}, \bibinfo {author}
  {\bibfnamefont {A.~R.}\ \bibnamefont {Akhmerov}}, \ and\ \bibinfo {author}
  {\bibfnamefont {C.~W.~J.}\ \bibnamefont {Beenakker}},\ }\href {\doibase
  10.1103/PhysRevB.88.035121} {\bibfield  {journal} {\bibinfo  {journal} {Phys.
  Rev. B}\ }\textbf {\bibinfo {volume} {88}},\ \bibinfo {pages} {035121}
  (\bibinfo {year} {2013})}\BibitemShut {NoStop}%
\bibitem [{\citenamefont {Dutta}\ \emph {et~al.}(2016)\citenamefont {Dutta},
  \citenamefont {Rahmani},\ and\ \citenamefont {del Campo}}]{dutta}%
  \BibitemOpen
  \bibfield  {author} {\bibinfo {author} {\bibfnamefont {A.}~\bibnamefont
  {Dutta}}, \bibinfo {author} {\bibfnamefont {A.}~\bibnamefont {Rahmani}}, \
  and\ \bibinfo {author} {\bibfnamefont {A.}~\bibnamefont {del Campo}},\ }\href
  {\doibase 10.1103/PhysRevLett.117.080402} {\bibfield  {journal} {\bibinfo
  {journal} {Phys. Rev. Lett.}\ }\textbf {\bibinfo {volume} {117}},\ \bibinfo
  {pages} {080402} (\bibinfo {year} {2016})}\BibitemShut {NoStop}%
\bibitem [{\citenamefont {Pontryagin}(1987)}]{pontryagin_mathematical_1987}%
  \BibitemOpen
  \bibfield  {author} {\bibinfo {author} {\bibfnamefont {L.~S.}\ \bibnamefont
  {Pontryagin}},\ }\href@noop {} {\emph {\bibinfo {title} {Mathematical Theory
  of Optimal Processes}}}\ (\bibinfo  {publisher} {{CRC} Press, Boca Raton,
  FL},\ \bibinfo {year} {1987})\BibitemShut {NoStop}%
\bibitem [{Note1()}]{Note1}%
  \BibitemOpen
  \bibinfo {note} {This special scenario does not occur in our
  system.}\BibitemShut {Stop}%
\bibitem [{\citenamefont {Karzig}\ \emph {et~al.}()\citenamefont {Karzig},
  \citenamefont {Oreg}, \citenamefont {Refael},\ and\ \citenamefont
  {Freedman}}]{karzig_geometric_2016}%
  \BibitemOpen
  \bibfield  {author} {\bibinfo {author} {\bibfnamefont {T.}~\bibnamefont
  {Karzig}}, \bibinfo {author} {\bibfnamefont {Y.}~\bibnamefont {Oreg}},
  \bibinfo {author} {\bibfnamefont {G.}~\bibnamefont {Refael}}, \ and\ \bibinfo
  {author} {\bibfnamefont {M.~H.}\ \bibnamefont {Freedman}},\ }\href@noop {}
  {}\Eprint {http://arxiv.org/abs/1511.05161} {arXiv:1511.05161} \BibitemShut
  {NoStop}%
\bibitem [{\citenamefont {Zhang}(2011)}]{zhang}%
  \BibitemOpen
  \bibfield  {author} {\bibinfo {author} {\bibfnamefont {F.}~\bibnamefont
  {Zhang}},\ }\href@noop {} {\emph {\bibinfo {title} {Matrix Theory: Basic
  Results and Techniques}}}\ (\bibinfo  {publisher} {Springer, Berlin},\
  \bibinfo {year} {2011})\BibitemShut {NoStop}%
\bibitem [{\citenamefont {Pichler}\ \emph {et~al.}(2013)\citenamefont
  {Pichler}, \citenamefont {Schachenmayer}, \citenamefont {Daley},\ and\
  \citenamefont {Zoller}}]{pilcher_heating_2013}%
  \BibitemOpen
  \bibfield  {author} {\bibinfo {author} {\bibfnamefont {H.}~\bibnamefont
  {Pichler}}, \bibinfo {author} {\bibfnamefont {J.}~\bibnamefont
  {Schachenmayer}}, \bibinfo {author} {\bibfnamefont {A.~J.}\ \bibnamefont
  {Daley}}, \ and\ \bibinfo {author} {\bibfnamefont {P.}~\bibnamefont
  {Zoller}},\ }\href {\doibase 10.1103/PhysRevA.87.033606} {\bibfield
  {journal} {\bibinfo  {journal} {Phys. Rev. A}\ }\textbf {\bibinfo {volume}
  {87}},\ \bibinfo {pages} {033606} (\bibinfo {year} {2013})}\BibitemShut
  {NoStop}%
\bibitem [{\citenamefont {Rahmani}(2015)}]{rahmani_dynamics_2015}%
  \BibitemOpen
  \bibfield  {author} {\bibinfo {author} {\bibfnamefont {A.}~\bibnamefont
  {Rahmani}},\ }\href {\doibase 10.1103/PhysRevA.92.042110} {\bibfield
  {journal} {\bibinfo  {journal} {Phys. Rev. A}\ }\textbf {\bibinfo {volume}
  {92}},\ \bibinfo {pages} {042110} (\bibinfo {year} {2015})}\BibitemShut
  {NoStop}%
\bibitem [{\citenamefont {D'Alessio}\ and\ \citenamefont
  {Rahmani}(2013)}]{Dalessio}%
  \BibitemOpen
  \bibfield  {author} {\bibinfo {author} {\bibfnamefont {L.}~\bibnamefont
  {D'Alessio}}\ and\ \bibinfo {author} {\bibfnamefont {A.}~\bibnamefont
  {Rahmani}},\ }\href {\doibase 10.1103/PhysRevB.87.174301} {\bibfield
  {journal} {\bibinfo  {journal} {Phys. Rev. B}\ }\textbf {\bibinfo {volume}
  {87}},\ \bibinfo {pages} {174301} (\bibinfo {year} {2013})}\BibitemShut
  {NoStop}%
\bibitem [{\citenamefont {Kasdin}(1995)}]{Kasdin}%
  \BibitemOpen
  \bibfield  {author} {\bibinfo {author} {\bibfnamefont {J.}~\bibnamefont
  {Kasdin}},\ }\href {https://doi.org/10.1109/5.381848} {\bibfield  {journal}
  {\bibinfo  {journal} {Proc. IEEE,}\ }\textbf {\bibinfo {volume} {83}},\
  \bibinfo {pages} {802} (\bibinfo {year} {1995})}\BibitemShut {NoStop}%
\end{thebibliography}%

\end{document}